\documentclass[11pt]{amsart}

\usepackage{amssymb}
\usepackage{amsmath,amsthm,amsfonts}  
 \usepackage{graphicx}
\usepackage{amssymb}
\usepackage{amsmath}
\usepackage{amsthm}
\usepackage{pdfsync}
\usepackage{fancyhdr}
\usepackage[applemac]{inputenc}
\usepackage{mathabx}

\usepackage[default]{frcursive}
\usepackage[T1]{fontenc}

\usepackage{color}
\usepackage{xcolor}
\usepackage[breaklinks,colorlinks,backref]{hyperref}
\hypersetup{
    colorlinks, %set true if you want colored links
    linktoc=all, %set to all if you want both sections and subsections linked
    linkcolor=red,  %choose some color if you want links to stand out
  citecolor=hyptxt,
  urlcolor=blue}
  \hypersetup{
  citebordercolor=,
  filebordercolor=green,
  linkbordercolor=blue
}
\definecolor{hyptxt}{rgb}{0.7, 0.4, 0.9}
\usepackage{bibtopic}

\usepackage[full]{textcomp} % to get the right copyright, etc.
     \usepackage{fbb} % osf for text, lining for math
     \usepackage[scaled=.95]{cabin}
     \usepackage[varqu,varl]{inconsolata}% typewriter
     \usepackage[libertine,bigdelims,vvarbb]{newtxmath}
     \usepackage[cal=boondoxo]{mathalfa}% less slanted than STIX
     \usepackage[T1]{fontenc}
\useosf
\definecolor{hervecolor}{rgb}{0.8,0,0.7}
     
\newcommand{\ket}[1]{|\kern.3ex#1\kern.3ex\rangle}
\newcommand{\bra}[1]{\langle\kern.3ex #1 \kern.3ex|}
\newcommand{\scalar}[2]{\langle\kern.3ex #1 \kern.3ex|\kern.3ex#2\kern.3ex\rangle}

\newcommand{\ii}{\mathsf{i}}

\def\R{\mathbb{R}}
\def\N{\mathbb{N}}
\def\C{\mathbb{C}}

\def\lg{\langle }
\def\rg{\rangle }

\def\adg{a^{\dag}}

\def\vk{\varkappa}

\def\ud{\mathrm{d}}

\def\sfX{\mathsf{X}}

\def\sfP{\mathsf{P}}

\def\hN{\hat N}
\def\adgh{\left(a^{\bsh}\right)^{\dag}}
\def\bsh{\boldsymbol{\mathsf{h}}}
\def\adgh{\left(a^{\bsh}\right)^{\dag}}
\numberwithin{equation}{section}

\begin{document}
\date{\today}
 
\title[CS in Quantum Optics]{Coherent states in Quantum Optics: An oriented overview}
\author[J.-P. Gazeau]{
Jean-Pierre Gazeau$^{\mathrm{a,b}}$}

\address{\emph{  $^{\mathrm{a}}$ APC, UMR 7164,}\\
\emph{Univ Paris  Diderot, Sorbonne Paris Cit\'e}  
\emph{75205 Paris, France}} 

\address{\emph{$^{\mathrm{b}}$ Centro Brasileiro de Pesquisas F\'{\i}sicas } \\
\emph{Rua Xavier Sigaud 150, 22290-180 - Rio de Janeiro, RJ, Brazil  }}

\email{e-mail:
gazeau@apc.in2p3.fr}

{\abstract{ In this survey, various generalisations of Glauber-Sudarshan coherent states are described in a unified way, with their statistical properties and their possible role in non-standard quantisations of the classical  electromagnetic field. Some statistical photon-counting aspects of Perelomov SU$(2)$ and SU$(1,1)$ coherent states are emphasized. }} 
\maketitle

\tableofcontents

\section{Introduction}
\label{intro}
The aim of this contribution is to give a restricted review on  coherent states in a wide sense (linear, non-linear, and various other types), and on their possible relevance to quantum optics, where they are generically denoted by $|\alpha\rg$, for a complex parameter $\alpha$, with $\vert \alpha\vert < R$, $R\in (0,\infty)$. Many important aspects of these states, understood here in a wide sense, will not be considered, like  photon-added, intelligent, squeezed, dressed, ``non-classical", all those cat superpositions of any type, involved into quantum entanglement and information, .... Of course, such a variety of features can be found in existing articles or reviews. A few of them \cite{kindaoud01,hach_etal16,cruzgress17,hohumaz18,gohoza12,hach_etal18} are  included in the list of references in order to provide the reader with an extended palette of various other references.  

We have attempted to give a minimal framework for all various families of  $|\alpha\rg$'s which are described in the present review.  Throughout the paper we put $\hbar = 1=c$, except if we need to make precise physical units. In Section \ref{fock} we recall the main characteristics of the Hilbertian framework (one-mode) Fock space with the underlying Weyl-Heisenberg algebra of its lowering and raising operators, and the basic statistical interpretation in terms of detection probability.  In  Section \ref{genset} we introduce coherent states in Fock space as superpositions of number states with coefficients depending on a complex number $\alpha$. These ``PHIN" states are requested to obey two fundamental properties, normalisation and resolution of the identity in Fock space. The physical meaning of  the parameter $\alpha$ is explained in terms of the number of photons, and may or not be interpreted  in terms of classical optics quadratures. A first example is given in terms of holomorphic Hermite polynomials. We  then define  an important subclass AN in PHIN.
Section \ref{GSCS} is devoted to the celebrated prototype of all CS in class AN, namely the Glauber-Sudarshan states. Their multiple properties are recalled, and their fundamental role in quantum optics is briefly described by following the seminal 1963 Glauber paper.  We end the section with a description of the CS issued from unitary displacement of an arbitrary number eigenstate in place of the vacuum. The latter belong to the PHIN class, but not in the AN class. The so-called non-linear CS in the AN class are presented in   Section \ref{NLCS}, and an  example of $q$-deformed CS illustrates this important extension of standard CS. 
In Section \ref{spinCS} we adapt the Gilmore-Perelomov spin or SU$(2)$ CS to the quantum optics framework and we emphasize their statistical meaning in terms of photon counting. We extend them also these CS  to those issued from an arbitrary number state. We follow a similar approach  in  Section \ref{SUIICS} with  Perelomov and Barut-Girardello SU$(1,1)$ CS. 
Section \ref{sussglo} is devoted to another type of AN CS, named  Susskind-Glogower, which reveal to be quite attractive in the context of quantum optics. We end in Section \ref{DFBCS} this list of various CS with a new type of non-linear CS based on deformed binomial distribution. 
In Section \ref{stat} we briefly review the statistical aspects of CS in quantum optics by focusing on their potential statistical properties, like sub- or super- Poissonian or just Poissonian.   The content of Section \ref{ANCSQ} concerns the role of all these generalizations of CS belonging to the AN class in the quantization of classical solutions of the Maxwell equations and the corresponding quadrature portraits. 
Some promising features of this CS quantization are discussed in Section \ref{conclu}.  

\section{Fock space}
\label{fock}
In their number or Fock representation,  the eigenstates  of the harmonic oscillator are simply denoted by kets $|n\rg$, where  $n=0, 1, \dotsc,$ stands for the number of elementary quanta of energy, named photons when the model is applied to a quantized monochromatic electromagnetic wave. These kets form an orthonormal basis of the Fock Hilbert space  $\mathcal{H}$. The latter is actually a  physical model  for all separable Hilbert spaces, namely the space $\ell^2(\N)$ of   square summable sequences.
For  such a basis (actually for any hilbertian basis $\{e_n\,,\, n= 0, 1, \dotsc\}$), the \emph{lowering} or \emph{annihilation}  operator  $a$, and its adjoint $a^{\dag}$, the \emph{raising} or \emph{creation} operator, are defined by
\begin{equation}
\label{acaadag}
a | n \rg = \sqrt{n} |n -1\rg\, , \quad \adg |n \rg = \sqrt{n+1} |n +
1\rg\, ,
\end{equation}
together with the action of $a$ on the ground or ``vacuum'' state 
$ a |0 \rg = 0$. They obey the so-called canonical commutation rule (ccr) $[a,\adg]= I$.
In this context, the \emph{number} operator $\hN = \adg a $ is diagonal in the  basis $\{|n\rg, \ n \in \N\}$, with spectrum $\N$: $\hN |n\rg = n | n \rg$.

\section{General setting for coherent states in a wide sense}
\label{genset}
\subsection{The PHIN Class}
 A large class of one-mode optical coherent states can be written as the following normalized superposition of photon number states,
 \begin{equation}
\label{csgen}
|\alpha\rg = \sum_{n=0}^{\infty} \phi_n(\alpha)|n\rg\, , 
\end{equation}
where the complex parameter $\alpha$ lies in some bounded or unbounded subset $\mathfrak{S}$ of $\C$. Its physical meaning will be discussed below in terms of detection probability. Note that  the adjective ``coherent" is used  in a generic sense and should not be understood in the restrictive sense it was given originally by Glauber \cite{glauber63-1}. 
 The complex-valued  functions $\alpha \mapsto \phi_n(\alpha)$, from which the name ``PHIN class",  obey the  two conditions
\begin{align}
\label{condnorm}
 1   &=    \sum_{n=0}^{\infty} \vert \phi_n(\alpha)\vert^2\, , \quad  \alpha \in \mathfrak{S}\, , \quad \mbox{(normalisation)} \\
\label{condortho}  \delta_{nn^{\prime}}  &=   \int_{\mathfrak{S}}\ud^2\alpha \,\mathfrak{w}\left(\alpha\right)\,\overline{ \phi_{n}(\alpha)}\, \phi_{n^{\prime}}(\alpha)\, , \quad \mbox{(orthonormality)}\, ,
\end{align}
where $\mathfrak{w}\left(\alpha\right)$ is a weight function,  with  support $\mathfrak{S}$ in $\C$. While Eq. \eqref{condnorm}  is necessary, Eq. \eqref{condortho} might be optional, except if we request resolution of the identity in the Fock Hilbert space spanned by the number states:
\begin{equation}
\label{resid}
 \int_{\mathfrak{S}}\ud^2\alpha\,\mathfrak{w}\left(\alpha\right)\, |\alpha\rg\lg\alpha| = I\, .
\end{equation}
A finite sum in \eqref{csgen} due to  $\phi_n= 0$ for all $n$ larger than a certain $n_{\mathrm{max}}$  may be considered in this study. 

If the orthonormality condition \eqref{condortho} is satisfied with a positive weight function, it allows us to interpret the map 
\begin{equation}
\label{alprob}
\alpha \mapsto \vert\phi_n(\alpha)\vert^2 \equiv \varpi_n(\alpha)
\end{equation}
as a probability distribution, with parameter $n$,  on the support $\mathfrak{S}$ of  $\mathfrak{w}$ in $\C$, equipped with the  measure $\mathfrak{w}\left(\alpha\right)\,\ud^2\alpha$.

On the other hand, the normalisation condition  \eqref{condnorm} allows to interpret the discrete map 
\begin{equation}
\label{probdet}
n \mapsto  \varpi_n(\alpha) 
\end{equation}
as a probability distribution on $\N$, with parameter $\alpha$, precisely the probability to detect $n$ photons when the quantum light is in the coherent state $|\alpha\rg$. The average value of the number operator 
\begin{equation}
\label{meanN}
\bar n =\bar n(\alpha) := \lg\alpha |\hN|\alpha\rg=\sum_{n=0}^{\infty}n\,\varpi_n(\alpha)
\end{equation}
 can be viewed as the intensity (or energy up to a physical factor like $\hbar \omega$) of the state $|\alpha\rg$ of the  quantum monochromatic radiation under consideration. An optical phase space associated with this radiation may be defined as  the image of the map 
 \begin{equation}
\label{mapaln}
\mathfrak{S} \ni \alpha \mapsto \xi_{\alpha}=\sqrt{\bar n(\alpha)}\,e^{\ii \arg \alpha}\in \C\, .
\end{equation}
A statistical interpretation of the original set $\mathfrak{S}$ is made possible if one can  invert the map \eqref{mapaln}. 
Two examples of such an inverse map will be given in Section \ref{spinCS} and Subsection \ref{persu11} respectively, with interesting statistical interpretations. 

\subsection{A first example of PHIN CS with holomorphic Hermite polynomials}

These coherent states were introduced in \cite{gasza11}. Given a real number $0<s<1$, the functions $\phi_{n;s}$ are defined as
\begin{equation}
\label{hen}
\phi_{n;s}(\alpha):= \frac{1}{\sqrt{b_n(s)\mathcal{N}_s(\alpha)}} \,e^{-\alpha^2/2}\, H_n(\alpha)\,,
\quad \alpha\in\mathbb C\, .
\end{equation}
 The non-holomorphic part lies in the expression of $\mathcal{N}_s$
 \begin{equation*}
%\label{normholcs1}
\mathcal{N}_s(\alpha) =\frac{s^{-1}-s}{2 \pi}\,
 e^{-s\,X^2 + s^{-1}\,Y^2}\, , \ \alpha= X+\ii Y\, .
\end{equation*}
 The constant $b_n(s)$ is given by
 \begin{equation*}% \label{beee}
   b_n(s)=\frac {\pi\sqrt s}{1-s}\biggl(2\frac{1+s}{1-s}\biggr)^n
n!\,.
   \end{equation*}
The function $H_n(\alpha)$ is the usual Hermite polynomial  of degree $n$ \cite{magnus66},  
 considered here as  an holomorphic polynomial in the complex variable $\alpha$.
The  corresponding normalised coherent states 
\begin{equation}
\label{csholherm}
|\alpha;s\rg = \sum_{n=0}^{\infty} \phi_{n;s}(\alpha)|n\rg\, , 
\end{equation}
solve the identity in $\mathcal{H}$, 
 \begin{equation}
\label{idholCS}
\frac{s^{-1}-s}{2\pi}\int_{\C} \ud^2\alpha
\,  |\alpha;s\rg\lg \alpha;s|=I\, .
\end{equation}
Thus, in the present case we have the constant weight $\mathfrak{w}\left(\alpha\right)= \frac{s^{-1}-s}{2\pi}$. This resolution of the identity results from the orthogonality  relations verified by the holomorphic Hermite polynomials in the complex plane:
  \begin{equation}
   \label{f1}
   \int_{\C}\ud
X\, \ud Y \, \overline{H_n(X+\ii Y)}\,H_{n^{\prime}}(X+\ii
Y)\exp\biggl[-(1-s)X^2-\biggl(\frac 1 s -1\biggr)Y^2\biggr] =b_n(s)\delta_{n n^{\prime}}\, .
   \end{equation}
   Note that the map $\alpha \mapsto \bar n(\alpha)= \sum_n n \left\vert e^{-\alpha^2/2}H_n(\alpha)\right\vert^2$
   is not rotationally invariant. 

\subsection{The AN Class}
Particularly convenient to manage and mostly encountered are coherent states $|\alpha\rg$  for which the functions $\phi_n$ factorise as
\begin{equation}
\label{phifact}
\phi_n(\alpha)= \alpha^n\,h_n(\vert\alpha\vert^2)\, , \quad \sum_{n=0}^{\infty} \vert\alpha\vert^{2n}\vert h_n(\alpha)\vert^2=1\, , \quad \vert \alpha\vert < R\, ,
\end{equation}
where $R$ can be finite or infinite.  All coherent states of the above type  lie in the so-called AN class (AN for ``$\alpha n$'').
Then, due to Fourier angular integration in \eqref{condortho},  the orthonormality condition holds if  there exists an isotropic weight function $w$ such that the $h_n$'s solve the following kind of moment problem on the interval $[0,R^2]$,
\begin{equation}
\label{mom}
\int_{0}^{R^2}\ud u\,w(u)\,u^n\vert h_n(u)\vert^2 = 1\,, \quad n\in \N\, . 
\end{equation}
This $w$ is related  to the above $\mathfrak{w}$ through
\begin{equation}
\label{wmfw}
\mathfrak{w}\left(\alpha\right)= \frac{w(\vert\alpha\vert^2)}{\pi}\,. 
\end{equation}
Note that the probability \eqref{probdet} to detect $n$ photons when the quantum light is in such a AN coherent state $|\alpha\rg$ is expressed as a function of  $u=\vert\alpha\vert^2$ only,
\begin{equation}
\label{probdetAN}
n\mapsto \varpi_n(\alpha)\equiv \sfP_n\left(u\right) = u^n\,\left(h_n(u)\right)^2\,. 
\end{equation}
Hence, the map $\alpha \mapsto \bar n$  is here rotationally invariant: $\bar n = \bar n (u)$. 
On the other hand, the  probability distribution on the interval $[0,R^2]$, for a detected  $n$, that CS $|\alpha\rg$ have classical intensity $u$ is given by
\begin{equation}
\label{probclAN}
u\mapsto \varpi_n(\alpha)\equiv \sfP_n\left(u\right) \,. 
\end{equation}

\section{Glauber-Sudarshan CS} 
\label{GSCS}
\subsection{Definition and properties}
They are the most popular, of course, among the AN families, and historically the first ones to appear in QED with Schwinger (1953) \cite{schwinger53}, and in quantum optics with the 1963 seminal papers by Glauber \cite{glauber63-1,glauber63-2,glauber63-3} and Sudarshan \cite{sudarshan63}. See also some key papers like \cite{mandel_wolf65,cahglau69,agawo70} for further developments in quantum optics and quantum  field theory. They were introduced  in  quantum mechanics by Schr\"{o}dinger (1926) \cite{schrodinger26} and  later by Klauder \cite{klauder60,klauder63-1,klauder63-2}. They correspond to the Gaussian
\begin{equation}
\label{gausscase}
h_n(u)= \dfrac{e^{-u}}{\sqrt{n!}} \, ,
\end{equation}
and read 
\begin{equation}
 \label{wavpackcs3}
| \alpha\, \rangle = e^{-\vert \alpha\vert^2/2} \sum_{n=0}^{\infty} \frac{\alpha^n}{\sqrt{n!}}\, |n \rg.
\end{equation}
Here, the parameter, i.e., the\textit{ amplitude}, $\alpha= X + \ii Y$ represents an element of the optical phase space. Its cartesian components $X$ and $Y$ in the Euclidean plane are called quadratures. In complete analogy with the harmonic oscillator model, the quantity $u=\vert \alpha\vert^2$ is considered as the classical \textit{intensity} or \textit{energy} of the coherent state $|\alpha\rg$. 
The corresponding detection distribution is the familiar Poisson distribution
\begin{equation}
\label{poissondist}
n\mapsto \sfP_n(u)= e^{-u}\,\frac{u^n}{n!}\,, 
\end{equation}
and the average value of the number  operator is just the intensity.
 \begin{equation}
\label{stcsavern}
\bar n(\alpha)= \vert \alpha\vert^2=u\,. 
\end{equation}
Hence, the detection distribution is written in terms of this  average value  as 
\begin{equation}
\label{stcsavern1}
\sfP_n(u)= e^{-\bar n}\,\frac{\bar n^n}{n!}\,. 
\end{equation}
From now on the states \eqref{wavpackcs3} will be called \textit{standard coherent states}. They are  called harmonic oscillator CS when we consider the $|n\rg$'s as eigenstates of the corresponding quantum hamiltonian $H_{\mathrm{osc}}= \left(P^2 + Q^2\right)/2= \hN + 1/2$ with $Q= \dfrac{a+ \adg}{\sqrt{2}}$ and $P= \dfrac{a- \adg}{\ii\sqrt{2}}$. They are exceptional in the sense that they obey the following long list of properties that give them, on their whole own,  a
strong status of uniqueness.
\begin{itemize}
{\it
\item[\bf P$_0$] The map} $\C \ni \alpha \rightarrow | \alpha \rangle \in \mathcal{H}$ {\it is
continuous.
\vskip 0.3cm
\item[\bf P$_1$] $| \alpha\rangle$ is  eigenvector of annihilation operator: $a |\alpha \rangle = \alpha | \alpha\rangle$.
\vskip 0.3cm
\item[\bf P$_2$] The CS family resolves the unity:}
$ \int_{\C} \frac{\ud^2\alpha}{\pi}\, 
|\alpha\rangle
\langle \alpha |  = I\,.$
\vskip 0.3cm
{\it
\item[\bf P$_3$] The CS saturate the Heisenberg inequality \index{Heisenberg!inequality}: $\Delta X\, \Delta Y= \Delta Q\, \Delta P= 1/2$.
\vskip 0.3cm
\item[\bf P$_4$] The CS family is temporally stable \index{Temporally stable}: $e^{-\ii H_{\mathrm{osc}} t} |\alpha \rangle = e^{-\ii t/2}
| e^{-\ii t}\alpha \rangle$.
\vskip 0.3cm
\item[\bf P$_5$] The mean value (or  ``lower symbol'' \index{Symbol!lower}) of the
Hamiltonian $H_{\mathrm{osc}}$ mimics the classical
relation energy-action:
$\check{H}_{\mathrm{osc}}(\alpha) := \langle \alpha | H_{\mathrm{osc}} |\alpha \rangle =  \vert \alpha \vert^2 + \frac{1}{2}$.
\vskip 0.3cm
\item[\bf P$_6$] The CS family is the orbit of the ground state under the action of the
Weyl \textit{displacement} operator: $| \alpha \rangle = e^{(\alpha a^{\dagger}-\bar{\alpha}a)}| 0
\rangle \equiv D(\alpha) | 0 \rangle $.
\vskip 0.3cm
\item[\bf P$_7$] The unitary Weyl-Heisenberg covariance follows from the above:}\\
\hspace*{20mm}$\mathcal{U}(s, \zeta) | \alpha \rangle
= e^{\ii (s+\mathrm{Im}(\zeta\bar{\alpha}))}| \alpha+\zeta \rangle$ where  $\mathcal{U}(s, \zeta) := e^{\ii s}\,D(\zeta)$.
\vskip 0.3cm
\item[\bf P$_8$] {\it From P$_2$ the coherent states provide a straightforward
quantization scheme:}\\
\hspace*{20mm}{\it  Function  $ f(\alpha)\rightarrow$ Operator $A_f=\int_{\C} \frac{\ud^2\alpha}{\pi}\,f(\alpha)\,|
\alpha\rangle
\langle \alpha |\,. $ }
\end{itemize}
\vskip 0.3cm
These properties cover a wide spectrum, starting from the
``wave-packet'' expression (\ref{wavpackcs3}) together with 
Properties  P$_3$ and P$_4$,  through an algebraic side (P$_1$),
a group representation side (P$_6$ and  P$_7$),
a functional analysis side (P$_2$) to end with the ubiquitous
problematic of
the relationship between classical and quantum models (P$_5$ and  P$_8$). Starting from this exceptional  palette of properties, the game over the past almost seven decades has been to build families of CS owing some of these properties, if not all of them, as it can be attested by the huge literature, articles, proceedings, special issues and author(s) or collective  books, a few of them being  \cite{perel72,KlauSkag85,perel86,zhangfenggil90,FengKlau94,aagbook14,dodonov02,dodoman03,vourdas06,gazeaubook09,alanbaga12,anbaga18}. 

\subsection{Why the adjective \emph{coherent}? (partially extracted from \cite{gazeaubook09})}

Let us compare the two  equations :
\begin{equation}
\label{compeig}
a |\alpha\rg = \alpha |\alpha \rg\, , \qquad a |n \rg = \sqrt{n}  |n -1 \rg\, .
\end{equation}
Hence,  \emph{an infinite superposition of number states  $|n\rg$, each of the latter describing a determinate number of elementary quanta, describes a state which is left unmodified (up to a factor) under the action of the  operator annihilating an elementary quantum. The factor is equal to the parameter $\alpha$ labeling the considered coherent state.}

More generally, we have  $f(a) |\alpha\rg = f(\alpha) |\alpha \rg$ for an analytic function $f$. This is precisely the idea developed by Glauber \cite{glauber63-1,glauber63-2,glauber63-3}.
Indeed, an electromagnetic field in a box can be assimilated to a countably infinite assembly of harmonic oscillators. This results from a simple Fourier analysis of Maxwell equations. The (canonical) quantization of these classical harmonic oscillators yields the Fock space  $\mathcal{F}$ spanned by all possible tensor products of  number eigenstates $ \bigotimes_k | n_k \rg \equiv | n_1, n_2, \dotsc , n_k, \dotsc\rg$, where 
 ``$k$'' is a shortening for  labeling  the mode \index{Mode} (including the photon polarization \index{Polarization})
 \begin{equation}
\label{emmode}
k \equiv \left\lbrace \begin{array}{cc}
   \vec{k}   &  \  \mbox{wave vector}\, , \\
      \omega_k  = \Vert \vec{k} \Vert c&   \  \mbox{frequency}\, ,\\
      \lambda =  1,2 & \ \mbox{helicity}\, , \ 
\end{array}\right. 
\end{equation}
and  $n_k$ is  the  number of photons in the   mode ``$k$''. 
 The Fourier expansion of the quantum vector potential   reads as 
 \begin{equation}
\label{FourexpA}
\overrightarrow{A}(\vec{r},t) = c \sum_k \sqrt{\frac{\hbar}{2 \omega_k}} \left( a_k \vec{u}_k (\vec{r}) e^{-\ii \omega_k t} +  a_k^{\dag} \overline{\vec{u}_k (\vec{r})} e^{\ii \omega_k t}\right)\, .
\end{equation} 
As an operator,   it acts (up to a gauge) on the Fock space   $\mathcal{F}$ \emph{via} $a_k$ and  $a_k^{\dag}$ defined by 
\begin{equation}
\label{actak}
a_{k_0} \prod_k | n_k \rg = \sqrt{n_{k_0}} | n_{k_0} -1\rg \prod_{k \neq k_0} | n_k \rg\, ,
\end{equation}
and obeying the canonical commutation rules
\begin{equation}
\label{ccrk}
 [ a_k, a_{k'}] = 0 =  [ a_k^{\dag}, a_{k'}^{\dag}]\, , \qquad   [ a_k, a_{k'}^{\dag}] = \delta_{kk'}\, I\, .
\end{equation}

Let us now give more insights on the modes, observables, and Hamiltonian. 
On the level of the  mode functions $\vec{u}_k$ the Maxwell equations read as:
\begin{equation}
\label{modeq}
 \Delta \vec{u}_k (\vec{r}) + \frac{\omega_k^2}{c^2} \vec{u}_k (\vec{r}) = \vec{0}\, .
\end{equation}
 When confined to a cubic box  $C_L$ with size $L$, these functions  form an  orthonormal basis 
 \begin{equation*}
\int_{C_L} \overline{\vec{u}_k (\vec{r})}\cdot \vec{u}_{l} (\vec{r})\, \ud^3 \vec{r} = \delta_{kl}\, ,
\end{equation*}
with obvious discretization constraints on ``$k$''.
By choosing the gauge $\nabla \cdot \vec{u}_k (\vec{r}) = 0,$ their expression is:  
\begin{equation}
\label{expmode}
\vec{u}_k (\vec{r}) =  L^{-3/2} \widehat{e}^{(\lambda)} e^{\ii \vec{k}\cdot \vec{r}}\, , \quad \lambda = 1 \ \mbox{or} \ 2\, , \quad  \vec{k}\cdot \widehat{e}^{(\lambda)} = 0\, , 
\end{equation}
where the $\widehat{e}^{(\lambda)}$'s stand for polarization vectors. 
The respective expressions of the electric  and magnetic field operators are  derived  from   the vector potential:
 \begin{equation*}
\overrightarrow{E} = - \frac{1}{c} \frac{\partial\overrightarrow{A}}{\partial t}\, , \quad  \overrightarrow{B}= \overrightarrow{\nabla} \times \overrightarrow{A}\, .
\end{equation*}
Finally, the electromagnetic field Hamiltonian is given by:
\begin{equation*}
H_{\mathrm{e.m.}} = \frac{1}{2}\int\left( \Vert \overrightarrow{E} \Vert^2 + \Vert \overrightarrow{B} \Vert^2\right)\, \ud^3\vec{r} = \frac{1}{2} \sum_k \hbar \omega_k \left( a_k^{\dag} a_k + a_k  a_k^{\dag}\right)\, .
\end{equation*}

Let us now decompose the electric field operator into positive and negative frequencies.
  \begin{align}
\label{pmfreq}
\nonumber  \overrightarrow{E}  & = \overrightarrow{E}^{(+)} + \overrightarrow{E}^{(-)}, \  \overrightarrow{E}^{(-)} = {\overrightarrow{E}^{(+)}}^{\dag}\, ,\\
    \overrightarrow{E}^{(+)} (\vec{r}, t)  & = \ii \sum_k \sqrt{\frac{\hbar \omega_k}{2}} a_k \vec{u}_k(\vec{r}) e^{-\ii \omega_k t}\,.
\end{align}
We then consider the field described by the density  (matrix) operator \index{Density!operator} \index{Density!matrix} :
\begin{equation}
\label{density}
\rho = \sum_{(n_k)} c_{(n_k)} \prod_{k} |n_k \rg\lg n_k |\,, \quad c_{(n_k)} \geq 0\, , \quad \mathrm{tr}\, \rho = 1\,  ,
\end{equation}
and the derived sequence of correlation functions \index{Correlation!function} 
$G^{(n)}$. The Euclidean tensor components  for the simplest one read as 
\begin{equation}
\label{compcorr}
G_{ij}^{(1)} (\vec{r},t; \vec{r^{\prime}},t^{\prime}) = \mathrm{tr}\left\{ \rho  E_i^{(-)} (\vec{r}, t) E_j^{(+)} (\vec{r^{\prime}}, t^{\prime})\right\}\,, \quad i,j = 1, 2, 3\,  .
\end{equation}
 They measure the correlation of the field state at different space-time points.
A \emph{coherent state} or  \emph{coherent radiation} \index{Coherent!radiation} $|\mbox{c.r.}\rg$ for the electromagnetic field is then defined by
\begin{equation}
\label{cohrad}
 |\mbox{c.r.}\rg = \prod_{k} | \alpha_k \rg \, ,
\end{equation}
where $| \alpha_k \rg $ is precisely the  standard coherent state  for the ``{$k$}" mode :
\begin{equation}
\label{cskmode}
| \alpha_k \rg = e^{-\frac{\vert \alpha_k \vert^2}{2}} \sum_{n_k} \frac{(\alpha_k)^{n_k}}{\sqrt{n_k!}} | n_k \rg\,, \quad a_k | \alpha_k \rg =  \alpha_k| \alpha_k \rg\,  ,
\end{equation}
with $\ \alpha_k \in \C $.   
The particular status of the state $|\mbox{c.r.}\rg$ is well understood through the action of the positive frequency electric field operator
 \begin{equation}
\label{actoncs}
 \overrightarrow{E}^{(+)} (\vec{r}, t) |\mbox{c.r.}\rg =  \overrightarrow{\mathcal{E}}^{(+)} (\vec{r}, t) |\mbox{c.r.}\rg\, .
\end{equation} 
The expression $\overrightarrow{\mathcal{E}}^{(+)} (\vec{r}, t)$ which shows up is precisely the classical field expression, solution to the Maxwell equations. 
\begin{equation}
\label{classexpr}
\overrightarrow{\mathcal{E}}^{(+)} (\vec{r}, t) = \ii \sum_k \sqrt{\frac{\hbar \omega_k}{2}} \alpha_k \vec{u}_k(\vec{r}) e^{-\ii \omega_k t}\, .
\end{equation}
Now, if the density operator is chosen as a pure coherent state, i.e.,
\begin{equation}
\label{purecs}
 \rho =  |\mbox{c.r.}\rg  \lg\mbox{c.r.}|\, , 
\end{equation}
  then the components (\ref{compcorr}) of the first order correlation function factorizes into independent terms  :
 \begin{equation}
\label{factcorr}
G_{ij}^{(1)} (\vec{r},t; \vec{r^{\prime}},t^{\prime}) = \overline{\mathcal{E}_i^{(-)} (\vec{r}, t)} \mathcal{E}_j^{(+)} (\vec{r^{\prime}}, t^{\prime})\, .
\end{equation}

\emph{An electromagnetic field operator is said ``fully coherent'' in the Glauber sense if  all of its correlation functions factorize like in (\ref{factcorr})}. Nevertheless, one should notice that such a definition does not imply monochromaticity.

A last important point concerns the production of such states in quantum optics. They can be manufactured by adiabatically coupling the e.m. field to a classical source, for instance a radiating current $\vec{j}(\vec{r}, t)$. The coupling is described by  the Hamiltonian 
\begin{equation}
\label{couplingH}
H_{\mathrm{coupling}} = -\frac{1}{c}\int\ud \vec{r} \,\overrightarrow{j}(\vec{r}, t)\cdot \overrightarrow{A}(\vec{r},t)\,. 
\end{equation}
From the Schr\"odinger equation, the time evolution of a field state  supposed to be originally, say at $t_0$, the state $|\mathrm{vacuum}\rg$(no photons) is given  by
\begin{equation}
\label{vactot}
|t\rg = \exp\left[\frac{\ii}{\hbar c}\int_{t_0}^{t}\ud t^{\prime}\int \ud \vec{r} \,\overrightarrow{j}(\vec{r}, t^{\prime})\cdot \overrightarrow{A}(\vec{r},t^{\prime}) + \ii \varphi(t)\right]\,|\mathrm{vacuum\rg}\, ,
\end{equation}
where $\varphi(t)$ is some phase factor, which cancels if one deals with the density  operator $|t\rg\lg t|$ and can be dropped.  From the Fourier expansion \eqref{FourexpA} we easily express the above evolution operator in terms 
of  the Weyl displacement  operators corresponding to each mode, 
\begin{equation}
\label{evopem}
\exp\left[\frac{\ii}{\hbar c}\int_{t_0}^{t}\ud t^{\prime}\int \ud \vec{r} \,\overrightarrow{j}(\vec{r}, t^{\prime})\cdot \overrightarrow{A}(\vec{r},t^{\prime}) \right] = \prod_{k}D(\alpha_k(t))\, , 
\end{equation}
where the complex amplitudes  are given by
\begin{equation}
\label{complampl}
\alpha_k(t) = \frac{\ii}{\hbar c}\int_{t_0}^{t}\ud t^{\prime}\int \ud \vec{r} \,\overrightarrow{j}(\vec{r}, t^{\prime})\cdot \overline{\vec{u}_k (\vec{r})} e^{\ii \omega_k t^{\prime}} \,. 
\end{equation}
Hence, we obtain the time-dependent e.m. CS
\begin{equation}
\label{emCSt}
|t\rg = \otimes_{k}|\alpha_k(t)\rg\,. 
\end{equation}
\subsection{Weyl-Heisenberg CS  with Laguerre polynomials}
The construction of the standard CS is minimal from the point of view of the  action of the Weyl unitary operator $D(\alpha)$ on the vacuum $|0\rg$ (Property \textbf{P}$_6$). More elaborate states, are issued from the action of  $D(\alpha)$ on other states $|s\rg$, $s=1,2, \dotsc$, of the Fock basis, which might be considered as initial states in the evolution described by \eqref{vactot}. Hence, let us define the family of CS 
\begin{equation}
\label{WHcss}
|\alpha;s\rg = D(\alpha)|s\rg = \sum_{n=0}^{\infty}D_{ns}(\alpha)|n\rg\,.
\end{equation}
The coefficients in this Fock expansion are the matrix elements $D_{ns} = \lg n|D(\alpha)|s\rg$ of the displacement operator. They are given in terms of the  generalized Laguerre polynomials \cite{magnus66} as
\begin{align}
\nonumber 
\label{Dlaguerre}
D_{ns}(\alpha) &:= \sqrt{\frac{s!}{n!}}\, e^{-\frac{\vert \alpha \vert ^2}{2}}\, \alpha^{n-s}\, L_{s}^{(n-s)}\left(\vert \alpha \vert^2\right)\quad \mbox{for}\quad s \leq n\, , \\
&= \sqrt{\frac{n!}{s!}}\, e^{-\frac{\vert \alpha \vert ^2}{2}}\, (-\bar{\alpha})^{s-n}\, L_{n}^{(s-n)}\left(\vert \alpha \vert^2\right)\quad \mbox{for}\quad s > n\, .
\end{align}
As matrix elements of a projective square-integrable UIR of the Weyl-Heisenberg group they obey the orthogonality relations
\begin{equation}
\label{orthmatel}
\int_{\C}\frac{\ud^2\alpha}{\pi}\, \overline{D_{ns}(\alpha})\,D_{n^{\prime}s^{\prime}}(\alpha)= \delta_{nn^{\prime}}\,\delta_{ss^{\prime}}\,.
\end{equation}
Like for the general case presented in \eqref{condortho}-\eqref{resid} this property validates the resolution of the identity
\begin{equation}
\label{residns}
 \int_{\C^2}\frac{\ud^2\alpha}{\pi}\, |\alpha;s\rg\lg\alpha;s| = I\, .
\end{equation}
The corresponding detection distribution is the ``Laguerre weighted" Poisson distribution,
\begin{equation}
\label{poissonw}
n\mapsto \sfP_n(u)= \left\lbrace \begin{array}{cc}
  e^{-u}\,\dfrac{u^{s-n}}{(s-n)!} \,\dfrac{\left(L_{n}^{(s-n)}(u)\right)^2 }{\binom{s}{n}}   &  n\leq s  \\
   e^{-u}\,\dfrac{u^{n-s}}{(n-s)!} \,\dfrac{\left(L_{s}^{(n-s)}(u)\right)^2 }{\binom{n}{s}}    &   n\geq s
\end{array}\right.\,.
\end{equation}
Of course, the optical phase space made of the complex $\sqrt{\bar n(\alpha)}e^{\ii\arg \alpha}$ is here less immediate. 

We notice that for $s>0$, these CS  $|\alpha;s\rg $  do not pertain to the AN class, since we find in the expansion a finite  number of terms in $\bar\alpha^n$ besides  an infinite number of terms in $\alpha^n$. 
On the other hand, there exist families of coherent states in the AN class (or their complex conjugate) which  are related to the  generalized Laguerre polynomials in a quasi-identical way \cite{cotgagor10,albaga12}.

\section{Non-linear CS}
\label{NLCS}
\subsection{General}
We define as non-linear CS those AN CS for which the functions $h_n(u)$ assume the simple form
\begin{equation}
\label{nlgpn}
h_n(u)= \frac{\lambda_n}{\sqrt{\mathcal{N}(u)}}\, , \quad \mathcal{N}(u)= \sum_{n=0}^{\infty} \vert \lambda_n\vert^2 u^n\, . 
\end{equation}
\subsection{Deformed Poissonian CS}
They are particular cases of the above. All $\lambda_n$  form a strictly decreasing sequence of positive numbers tending to $0$:
\begin{equation}
\label{defpoiss}
\lambda_0 = 1 > \lambda_1 > \cdots \lambda_n > \lambda_{n+1} > \cdots\, , \quad \lambda_n \to 0\,.  
\end{equation}
We now introduce the strictly increasing sequence,
\begin{equation}
\label{lnxn}
x_n = \left(\frac{\lambda_{n-1}}{\lambda_n}\right)^2\, , \quad x_0= 0\, . 
\end{equation}
It is straightforward to check that
\begin{equation}
\label{lnxnf}
\lambda_n = \frac{1}{\sqrt{x_n!}}\, , \quad \mbox{with}\quad x_n! := x_1x_2\cdots x_n\, .
\end{equation}
Then $\mathcal{N}(u)$ is the generalized exponential with convergence radius $R^2$,
\begin{equation}
\label{genexp}
\mathcal{N}(u) = \sum_{n=0}^{\infty}  \frac{u^n}{x_n!}\, ,
\end{equation}
and the corresponding CS take the form extending to the non-linear case the familiar Glauber-Sudarshan one,
\begin{equation}
\label{DPNLCS}
|\alpha\rg = \frac{1}{\sqrt{\mathcal{N}(\vert\alpha)\vert^2)}}\sum_{n=0}^{\infty} \frac{\alpha^n}{\sqrt{x_n!}}|n\rg\, . 
\end{equation}
The orthonormality condition \eqref{condortho} is completely fulfilled if there exists a weight $w(u)$ solving   the  moment problem for the sequence $(x_n!)_{n\in \N}$, 
\begin{equation}
\label{mompbxn}
x_n!= \int_0^{R^2}\ud u\,\frac{w(u)}{\mathcal{N}(u)}\, u^n\,. 
\end{equation}
The detection probability distribution is the deformed Poisson distribution: 
\begin{equation}
\label{poissondef}
n\mapsto \sfP_n(u)= \frac{1}{\mathcal{N}(u)}\,\frac{u^n}{x_n!}\,. 
\end{equation}
The average value of the number operator $\bar n$ is given by
\begin{equation}
\label{nlcsavern}
\bar n \left(\vert\alpha\vert^2\right))= \lg \alpha|\hN|\alpha\rg= u\,\left.\frac{\ud \log\mathcal{N}(u)}{\ud u}\right\vert_{u=\vert\alpha\vert^2}\,. 
\end{equation}

\subsection{Example with $q$ deformations of integers}
These  coherent states have been studied by many authors, see \cite{gazolmo13}, that we follow here, and references therein. They are built from the symmetric or bosonic $q$-deformation of natural numbers:
\begin{equation}
\label{qsdef}
x_n = {}^{[s]}[n]_{q} = \frac{q^n - q^{-n}}{q-q^{-1}} = {}^{[s]}[n]_{q^{-1}}\, , \quad q >0\,. 
\end{equation}
\begin{equation} 
\label{qxncs}
|\alpha\rg _q =  \frac{1}{\sqrt{\mathcal{N}_q(\vert \alpha \vert^2)}}\sum_{n=0}^{\infty}\frac{\alpha^n}{\sqrt{{}^{[s]}[n ]_{q}!}}\,  |n\rg\, ,
\end{equation}
where  its associated exponential is one of the so-called $q$ exponentials \cite{solkac}
\begin{equation}
\label{seqexp1}
\mathcal{N}_q(u)=\mathfrak{e}_q(u) \equiv  = \sum_{n = 0}^{+\infty} \frac{u^n}{{}^{[s]}[n ]_{q}!}\,. 
\end{equation}
This series defines the analytic entire function $\mathfrak{e}_q(z)$ in the complex plane for any positive $q$.
The CS $|\alpha\rg _q $ in the limit $q\to 1$ goes to the standard CS  $|\alpha\rg $.   The solution to the moment problem \eqref{mom} for $0<q<1$   is given by
\begin{equation*}
\label{momsolxn}
  \int_0^{\infty}  \ud u\, w_q(u)\,  \frac{u^n}{\mathfrak{e}_q(u) \,{}^{[s]}[n]_{q}!}= 1
\end{equation*}
with positive density  
\begin{equation*}   
w_q(t) = (q^{-1}-q)\sum_{j=0}^\infty  \, g_q\left(t\, \frac{q^{-1}-q}{q^{2j}}\right) \mathfrak{E}_q\left(-\frac{q^{2j}}{q^{-1}-q}\right) \, .
\end{equation*}
The function $g_q$ is given by 
\begin{equation*}
\label{momqnn}
g_q(u) = \frac{1}{\sqrt{2 \pi\, \vert \ln{q}\vert}}\, \exp\left\lbrack - 
\frac{\left\lbrack\ln \left(\frac{u}{\sqrt{q}}\right)\right\rbrack^2}{2 \vert \ln q\vert}\right\rbrack\, .
\end{equation*}
and a second $q$-exponential \cite{solkac} appears here,
\begin{equation*}
\label{seqexp2}
 \mathfrak{E}_q (u) := \sum_{n=0}^{\infty} q^{\frac{n(n+1)}{2}}\frac{u^n}{{}^{[s]}[n]_{q}!} \, . 
\end{equation*}
Its radius of convergence is $\infty$ for $0< q \leq 1$ (it is equal to $1/(q-q^{-1})$ for $q> 1$). There results the resolution of the identity,
\begin{equation}
\label{qresid}
 \int_{\C}\ud^2\alpha\,\mathfrak{w}_q\left(\alpha\right)\, |\alpha\rg_q{}_q\lg\alpha| = I\, , \quad   \mathfrak{w}_q\left(\alpha\right)= \frac{w_q(\vert\alpha\vert^2)}{\pi}\,.
\end{equation}
More exotic families of non-linear CS are, for instance, presented in \cite{bafregaha12}.

\section{Spin CS as optical CS}
\label{spinCS}
These states are an adaptation to the quantum optical context of the well-known Gilmore or Perelomov SU$(2)$-CS, also called spin CS \cite{perel72,perel86}. The Fock  space reduces to the finite-dimensional  subspace $\mathcal{H}_j$, with dimension $n_j+1:=2j+1$, for $j$ positive integer or half-integer, consistenly with the fact that  the functions $h_n$,  given here by
\begin{equation}
\label{binomialcase}
h_n(u)= \sqrt{\binom{n_j}{n}}\, (1 + u)^{-\frac{n_j}{2}}\,, \quad \binom{n_j}{n}=\frac{n_j!}{n!(n_j-n)!}\, . 
\end{equation}
cancel for $n> n_j$. 
%One might consider this ``maximal" number of photons as imposed by reasonable physical considerations, e.g., imposed by feasibility of optical instruments or light sources.  
The corresponding spin CS read
\begin{equation}
\label{zetaspincs}
|\alpha;n_j\rg =  \left(1 + \vert \alpha \vert^2\right)^{-\frac{n_j}{2}}\sum_{n=0}^{n_j} \sqrt{\binom{n_j}{n}}\, \alpha^{n}\, |n\rg\,.
\end{equation}
They resolve the unity in $\mathcal{H}_{n_j}$ in the following way:
\begin{equation}
\label{csresunzeta}
\frac{n_j+1}{\pi}\, \int_{\C}\frac{\ud^2\alpha}{(1 + \vert \alpha\vert^2)^2}\, |\alpha;n_j\rg \lg \alpha;n_j|  = I\,.
\end{equation}
The  detection probability distribution is  binomial: 
\begin{equation}
\label{binomialdist}
n\mapsto \sfP_n(u)= (1+u)^{-n_j}\,\binom{n_j}{n}\,u^n\,. 
\end{equation}
There results the average value of the number operator
\begin{equation}
\label{avernphspin}
\bar n (u)= n_j\,\frac{u}{1+u}\ \Leftrightarrow\ u= \frac{\bar n/n_j}{1-\bar n/n_j} \,. 
\end{equation}
Thus the probability \eqref{binomialdist} is expressed in terms of the ratio $p:=\bar n/n_j$ as 
 \begin{equation}
\label{binomialdist1}
\sfP_n(u) \equiv \widetilde{\sfP}_n(p)= \binom{n_j}{n}\,(1-p)^{n_j-n} \, p^n\, ,
\end{equation}
which allows to define the optical phase space as the open disk of radius $\sqrt{n_j}$, $\mathcal{D}_{\sqrt{n_j}}= \left\{\xi_{\alpha}= \sqrt{\bar n\left(\vert \alpha\vert^2\right)}e^{\ii \arg \alpha}\, ,\,  \vert \xi_{\alpha}\vert < \sqrt{n_j}\right\}$. 

The interpretation of $\sfP_n(u)$ together with the number $n_j$ in terms of photon statistics (see Section \ref{stat} for more details) is luminous if we consider
a  beam of perfectly coherent light with a constant intensity. If the beam is of  finite length $L$ and is subdivided into $n_j$ segments of length $L/n_j$, then $\widetilde{\sfP}_n(p)$ is the probability of finding  $n$ subsegments containing one photon and $(n_j - n)$ containing no photons, in any possible order \cite{fox06}.   
A more general statistical  interpretation of \eqref{binomialdist} or \eqref{binomialdist1} is discussed in \cite{algahel08}. 

Note that the standard coherent states
 are  obtained from the above  CS at the limit $n_j \to \infty$  through a contraction process. The latter is carried out through a scaling of the complex variable $\alpha$, namely
 $\alpha \mapsto\sqrt{n_j}\, \alpha$. Then the binomial distribution $\widetilde{\sfP}_n(p)$ becomes the Poissionian
 \eqref{stcsavern1}, as expected. 
 
 Actually, these states are the simplest ones among a whole family issued from the Perelomov construction \cite{perel72,gazeaubook09,gahulare07}, and based on spin spherical harmonics. For our present purpose we modify their definition by including an extra phase factor and delete the factor $\sqrt{ \frac{2j+1}{4\pi}} $. For $j\in \N/2$ and  a given $-j\leq \sigma\leq j$,  the spin spherical harmonics  are the following functions  on the unit sphere $\mathbb{S}^2$. 
   \begin{align}
\label{jacspin}
    \nonumber _{\sigma}\mathfrak{Y}_{j \mu}(\Omega) &:= (-1)^{(j-\mu)}\sqrt{\frac{(j-\mu)!(j+\mu)!}{(j-\sigma)!(j+\sigma)!}}\times \\
\times&  \frac{1}{2^{\mu}}\, (1 + \cos{\theta})^{\frac{\mu + \sigma}{2}}\, (1 - \cos{\theta})^{\frac{\mu - \sigma}{2}}  P_{j-\mu}^{(\mu - \sigma,\mu + \sigma)}(\cos{\theta}) \,e^{-\ii (j-\mu) \varphi}\, , 
\end{align}
where $\Omega= (\theta,\varphi)$ (polar coordinates), $-j\leq \mu \leq j$, and the $P^{(a, b)}_{n} (x)$ are Jacobi polynomials \cite{magnus66} with $P^{(a, b)}_{0} (x) = 1$.  
Singularities of the factors at $\theta = 0$ (resp. $\theta = \pi$) for the power $ \mu - \sigma< 0$  (resp. $\mu + \sigma < 0$) are just apparent. To remove them  it is necessary to use alternate expressions of the Jacobi polynomials based on the relations:
\begin{equation}
\label{SpecJac}
P^{(-a, b)}_{n} (x) = \frac{\binom{n + b}{a}}{\binom{n }{a}}\left(\frac{x-1}{2}\right)^a P^{(a, b)}_{n-a} (x)\, . 
\end{equation}
The functions \eqref{jacspin} obeys the two conditions required in the construction of coherent states
\begin{align}
\label{orhtosph}
 \frac{2j +1}{4\pi}  \int_{\mathbb{S}^2}\ud \Omega\,  \overline{_{\sigma}\mathfrak{Y}_{j \mu}(\Omega)}\, _{\sigma}\mathfrak{Y}_{j \mu^{\prime}}(\Omega) &= \delta_{\mu \mu^{\prime}}   \quad \mbox{(orthogonality)}\, \\
  \label{normsph}  \sum_{\mu=-j}^{j} \vert_{\sigma}\mathfrak{Y}_{j \mu}(\Omega)\vert^2 &= 1   \quad \mbox{(normalisation)} \,.   
\end{align}
At $j=l$ integer and $\sigma =0$, $\mu = m$ we recover the spherical harmonics  $Y_{lm}(\Omega)$ (up to the factor $(-1)^le^{-\ii j \varphi} \sqrt{ \frac{2l+1}{4\pi}}$). 
 We now consider the parameter $\alpha$ in \eqref{zetaspincs}  
as issued from the stereographic projection  $\mathbb{S}^2 \ni\Omega \mapsto \alpha \in \C$:
\begin{equation}
\label{thetphial}
\alpha = \tan\frac{\theta}{2}\,e^{\ii \varphi}\, , \quad \mbox{with}\quad \ud \Omega= \sin\theta \ud \theta\ud\varphi= \frac{4\ud^2\alpha}{(1+\vert\alpha \vert^2)^2}\, . 
\end{equation}
In this regard, the probability $p=\bar n/n_j$ is equal to $\sin\theta/2$ while $\varphi=\arg \alpha$. 
With  the notations $n_j=2j\in \N$, $n=j-\mu= 0,1,2, \dotsc, n_j$, $0\leq s=j-\sigma\leq n_j$, adapted to the content of the present paper, and from the expression of the Jacobi polynomials, we get the functions  \eqref{jacspin} in terms of $\alpha\in \C$:
 \begin{equation}
\label{spinsalpha1}
_{\sigma}\mathfrak{Y}_{j \mu}(\Omega)= \alpha^n\, h_{n;s}\left(\vert\alpha\vert^2\right)\, , 
\end{equation}  
 where 
 \begin{equation}
\label{spinsalpha2}
h_{n;s}(u)= \sqrt{\frac{n!(n_j-n)!}{s!(n_j-s)!}}\,\left(1+u\right)^{-\frac{n_j}{2}}\sum_{r=\max(0,n+s-n_j)}^{\min(n,s)}\binom{s}{r}\,\binom{n_j-s}{n-r}\,(-1)^r\,u^{s/2-r}\,. 
\end{equation}
The corresponding ``Jacobi" CS are in the  AN class and read
\begin{equation}
\label{jacobiCS}
|\alpha;n_j;s\rg= \sum_{n=0}^{n_j} \alpha^n\, h_{n;s}\left(\vert\alpha\vert^2\right)\,|n\rg\,. 
\end{equation} 
They solve the identity as
\begin{equation}
\label{csnjsres}
\frac{n_j+1}{\pi}\, \int_{\C}\frac{\ud^2\alpha}{(1 + \vert \alpha\vert^2)^2}\, |\alpha;n_j;s\rg \lg \alpha;n_j;s|  = I\
\end{equation} 
 The states \eqref{zetaspincs} are recovered for $s=0$. Similarly to CS \eqref{WHcss} states \eqref{jacobiCS} can be also viewed as displaced occupied states. Indeed, they can be written in the Perelomov way  as
 \begin{equation}
\label{}
|\alpha;n_j;s\rg= \mathcal{D}^{n_j/2}\left(\zeta_{\alpha}\right)|s\rg\,,
\end{equation}
where $\zeta_{\alpha}= \begin{pmatrix}
   \left(1+ \vert \alpha\vert^2\right)^{-1/2}   & \left(1+ \vert \alpha\vert^2\right)^{-1/2}\,\alpha  \\
   -\left(1+ \vert \alpha\vert^2\right)^{-1/2}\,\bar\alpha   &   \left(1+ \vert \alpha\vert^2\right)^{-1/2}
\end{pmatrix}$  is the element of SU$(2)$ which brings $0$ to $\alpha$ under the homographic action $$\alpha \mapsto\begin{pmatrix}
   a   &  b  \\
    -\bar b  &  \bar a
\end{pmatrix}\cdot \alpha:= \dfrac{a\alpha+ b}{-\bar b \alpha + \bar a}$$ of this group on the complex plane, and  $\mathcal{D}^{n_j/2}$ is the corresponding $n_j+ 1$-dimensional UIR of SU$(2)$. Let us write   $\mathcal{D}^{n_j/2}\left(\zeta_{\alpha}\right)$ as a displacement operator similar to the Weyl-Heisenberg one (propriety \textbf{P}$_6$) and involving the usual angular momentum generators $J_{\pm}$ for the representation $\mathcal{D}^{n_j/2}$, 
\begin{equation}
\label{SU2disp}
\mathcal{D}^{n_j/2}\left(\zeta_{\alpha}\right)= e^{\varsigma_{\alpha} J_+ -\bar \varsigma_{\alpha} J_-}\equiv D_{n_j}(\varsigma_{\alpha})\, , \quad \varsigma_{\alpha}= -\tan^{-1}\vert\alpha\vert\,e^{-\ii \arg\alpha}\,. 
\end{equation}
Note that we could have adopted here the historical approaches by Jordan,  Holstein, Primakoff, Schwinger \cite{jordan35,holsprim40,schwinger52} in transforming these angular momentum operators in terms of  ``bosonic'' $a$ and $\adg$. Nevertheless this  QFT artificial flavour is not really useful in the present context. 

\section{SU$(1,1)$-CS as optical CS}
\label{SUIICS}

\subsection{Perelomov CS}
 \label{persu11}
These states are also an adaptation to the quantum optical context of the Perelomov SU$(1,1)$-CS \cite{perel72,perel86,gazeaubook09,gazolmo18}. They are yielded through a SU$(1,1)$ unitary action on a number state. The Fock Hilbert space $\mathcal{H}$ is infinite-dimensional while the complex number $\alpha$ is restricted to the open unit disk $\mathcal{D}:= \{\alpha  \in \C\, , \, \vert \alpha \vert < 1\}$. 
Let $\varkappa> 1/2$ and $s\in \N$. We then define the $(\varkappa;s)$-dependent CS family as  the ``SU$(1,1)$-displaced  $s$-th state'' 
\begin{equation}
\label{morecssu11}
|\alpha; \varkappa; s\rg=  U^{\varkappa}(p( \bar \alpha))|s\rg= \sum_{n = 0}^{\infty} U^{\varkappa}_{ns}(p( \bar \alpha)) |n\rg\equiv \sum_{n = 0}^{\infty} \phi_{n;\varkappa; s}(\alpha)\, |n\rg\, ,
\end{equation}
where the $U^{\varkappa}_{ns}(p( \bar \alpha))$'s are matrix elements of the  UIR $U^{\varkappa}$ of SU$(1,1)$ in its discrete series and $p( \bar\alpha)$ is the particular 
matrix 
\begin{equation}
\label{palapha}
\begin{pmatrix}
  \left(1-\vert \alpha\vert^2\right)^{-1/2}    &   \left(1-\vert \alpha\vert^2\right)^{-1/2}\,\bar\alpha   \\
   \left(1-\vert \alpha\vert^2\right)^{-1/2} \,\alpha    &   \left(1-\vert \alpha\vert^2\right)^{-1/2} 
\end{pmatrix}\in \mathrm{SU}(1,1)\,. 
\end{equation}
They are given in terms of Jacobi polynomials as
\begin{align}
\label{Upz}
\nonumber   U^{\varkappa}_{ns}(p(\bar \alpha)) & = \left( \frac{n_<!\, \Gamma(2 \varkappa + n_>) }{n_>!\, \Gamma(2 \varkappa + n_<)} \right)^{1/2}  \left(1-\vert \alpha \vert^2\right)^{\varkappa} \, (\mathrm{sgn}(n-s))^{n-s}\,\times \\
& \times   P^{(n_> - n_<\, ,\, 2 \varkappa -1)}_{n_<}\left( 1-2\vert \alpha\vert^2 \right)\,\times\left\lbrace\begin{array}{cc}
\alpha^{n-s}      & \mathrm{if}\ n_{>} =n  \\
 \bar\alpha^{s-n}     &   \mathrm{if}\ n_{>} =s
\end{array}\right.
\end{align}
with  $n_{\substack{
>\\
<}}  = \left\lbrace \begin{array}{c}
    \max      \\
       \min
\end{array}\right.\, (n,s) \geq 0$.
The states \eqref{morecssu11} solve the identity:
\begin{equation}
\label{runitmorecssu11}
\frac{2\varkappa-1}{\pi}\int_{\mathcal{D}}\frac{\ud^2 \alpha}{\left(1-\vert\alpha\vert^2\right)^2} \, |\alpha; \varkappa;s \rg \lg \alpha; \varkappa;s | = I\,.
\end{equation}

The simplest case $s=0$ pertains to the AN class,
\begin{equation}
\label{cssu11}
|\alpha;\varkappa;0\rg\equiv |\alpha;\varkappa\rg= \sum_{n=0}^{\infty}\alpha^n\,h_{n;\varkappa}\left(\vert\alpha\vert^2\right)\,|n\rg\, , \quad h_{n;\varkappa}(u):=  \sqrt{\binom{2\varkappa -1 +n}{n}}\, (1-u)^{\varkappa}\,.
\end{equation}
The corresponding  detection probability distribution is negative binomial,  
\begin{equation}
\label{negbinomialdist}
n\mapsto \sfP_n(u)= (1-u)^{2\varkappa}\,\binom{2\varkappa -1 +n}{n}\,u^n\,. 
\end{equation}
The average value of the number operator reads as
\begin{equation}
\label{avernphperel}
\bar n(u)= 2\varkappa\,\frac{u}{1-u}\ \Leftrightarrow\ u= \frac{\bar n/2\vk}{1+\bar n/2\vk} \,. 
\end{equation}
By introducing the ``efficiency'' $\eta := 1/2\vk\in (0,1)$ the probability \eqref{negbinomialdist} is expressed in terms of the corrected average value $\bar N:=\eta\bar n$ as 
 \begin{equation}
\label{negbinomialdist1}
\sfP_n(u) \equiv \widetilde{\sfP}_n(\bar N)= (1+\bar N)^{-1/\eta}\binom{1/\eta-1 +n}{n}\, \left(\frac{\bar N}{1+\bar N}\right)^n\, .
\end{equation}
It is remarkable that such a distribution reduces to the celebrated Bose-Einstein one for the thermal light at the limit $\eta= 1$, i.e., at the lowest bound $\vk = 1/2$ of the discrete series of SU$(1,1)$. For $\eta< 1$, the difference might be understood from the fact that we consider the average photocount number $\bar N$ instead of the mean photon number $\bar n$ impinging on the detector in the same interval
\cite{fox06}.   For a related interpretation within the framework of thermal
equilibrium states of the oscillator see \cite{aharonov73}
 
Note that the above CS, built from the  negative binomial distribution, were also discussed in \cite{algahel08}. 

Like for CS \eqref{WHcss},  the CS  $|\alpha;\varkappa;s\rg $ in \eqref{morecssu11} do not pertain to the AN class for $s>0$. In their expansion there are $s$ terms in $\bar\alpha^{s-n}$, $s>n$,  besides  an infinite number of terms in $\alpha^{n-s}$, $s\leq n$. Finally, like for the Weyl-Heisenberg and SU$(2)$ cases, the  representation operator $U^{\varkappa}(p( \bar \alpha))$ used in \eqref{morecssu11} to build the SU$(1,1)$ CS can  be given the following form of a displacement operator involving the  generators $K_{\pm}$ for the representation $U^{\kappa}$ \cite{perel86},  
\begin{equation}
\label{SU11disp}
U^{\kappa} (p(\bar \alpha)) = e^{\varrho_{\alpha}\, K_+ - \bar\varrho_{\alpha} \, K_-}\equiv D_{\kappa}(\varrho_{\alpha})\, , \quad \varrho_{\alpha}= \tanh^{-1}\vert \alpha\vert\,e^{\ii \arg \alpha}\, .  
\end{equation}

\subsection{Barut-Girardello CS}
These nonlinear CS states \cite{bargir71,angaklamopen01} pertain to the AN class. They are requested to be  eigenstates of the SU$(1,1)$ lowering operator in its discrete series representation $U^{\varkappa}$, $\varkappa> 1/2$. The Fock Hilbert space $\mathcal{H}$ is infinite-dimensional while the complex number $\alpha$  has no domain restriction in $\C$. With the notations of \eqref{DPNLCS} they read
\begin{equation}
\label{bargirCS}
|\alpha;\varkappa\rg_{\mathrm{BG}}= \frac{1}{\sqrt{\mathcal{N}_{\mathrm{BG}}(\vert\alpha\vert^2)}}\sum_{n=0}^{\infty} \frac{\alpha^n}{\sqrt{x_n!}}|n\rg\, , \quad x_n= n(2\varkappa +n-1)\, , \quad x_n!= n!\frac{\Gamma(2\varkappa + n)}{\Gamma(2\varkappa)}\, ,
\end{equation}
with
\begin{equation}
\label{ BGbessel}
\mathcal{N}_{\mathrm{BG}}(u)= \Gamma(2\varkappa)\sum_{n=0}^{\infty}\frac{u^n}{n!\Gamma(2\varkappa +n)}= \Gamma(2\varkappa)\,u^{-\varkappa}\,I_{2\varkappa-1}(2\sqrt{u}),
\end{equation}
where $I_{\nu}$ is a modified Bessel function \cite{magnus66}. In the present case the moment problem \eqref{mom} is solved as
\begin{equation}
\label{momBG}
\int_{0}^{\infty}\ud u\,w_{\mathrm{BG}}(u)\,\frac{u^n}{\mathcal{N}_{\mathrm{BG}}(u)\,x_n!}\, = 1\,, \quad w_{\mathrm{BG}}(u)= \mathcal{N}_{\mathrm{BG}}(u)\,\frac{2}{\Gamma(2\varkappa)}\,u^{\varkappa-1/2}\,K_{2\varkappa-1}(2\sqrt{u})\, ,
\end{equation}
where $K_\nu$ is the second modified Bessel function. The resolution of the identity follows:
\begin{equation}
\label{BGresunit}
\int_{\C} \ud^2 \alpha\,\mathfrak{w}_{\mathrm{BG}}\left(\alpha\right)\, |\alpha;\varkappa\rg_{\mathrm{BG}}{}_{\mathrm{BG}}\lg\alpha;\varkappa| = I\, ,\quad \mathfrak{w}_{\mathrm{BG}}(u)= \frac{w_{\mathrm{BG}}(u)}{\pi}\,. 
\end{equation}

\section{Adapted Susskind-Glogower  CS}
\label{sussglo}
Let us examine the Susskind-Glogower CS \cite{sussglog64} presented in \cite{moyasoto11}. These normalised states read for real $\alpha \equiv x\in \R$, 
\begin{equation}
\label{SGmoya}
|x\rg_{\mathrm{SG}}= \sum_{n=0}^{\infty}  (n+1)\,\frac{J_{n+1}(2x)}{x}\,|n\rg\,, 
\end{equation}
where the Bessel function $J_\nu$ is given by
\begin{equation}
\label{bessel}
J_\nu(z)= \left(\frac{z}{2}\right)^{\nu}\,\sum_{m=0}^{\infty}\frac{(-1)^m \left(\frac{z}{2}\right)^{2m}}{m!\,\Gamma(\nu+m+1)}\, . 
\end{equation} 
The normalisation implies the interesting identity \cite{evaldo-6-18}
\begin{equation}
\label{normbessel}
\sum_{n=1}^{\infty}  n^2\,\left(J_{n}(2x)\right)^2 = x^2\, .
\end{equation}
The above expression allows us to extend  the formula \eqref{SGmoya} in a non-analytic way to complex $\alpha$ as
\begin{equation}
\label{extendRSG}
(n+1)\frac{J_{n+1}(2x)}{x} \mapsto \alpha^n \,(n+1)\sum_{m=0}^{\infty}\frac{(-1)^m \vert \alpha\vert^{2m}}{m!\,\Gamma(n+m+2)}\equiv \alpha^n\,h^{\mathrm{SG}}_n(\vert\alpha\vert^2)\,,  
\end{equation}
i.e., \begin{equation}
\label{vpnbessel}
h^{\mathrm{SG}}_n(u) = (n+1)\,\frac{1}{u^{\frac{n+1}{2}}}\,J_{n+1}(2\sqrt{u})\, , 
\end{equation}
and thus
\begin{equation}
\label{CSSG}
|\alpha\rg_{\mathrm{SG}}= \sum_{n=0}^{\infty} \alpha^n\, h^{\mathrm{SG}}_n(\vert\alpha\vert^2)\,|n\rg\,. 
\end{equation}
The moment equation \eqref{mom} reads here 
\begin{equation}
\label{SGmom}
\int_{0}^{\infty}\ud u\,\frac{w(u)}{u}\,\left(J_{n}(2\sqrt{u})\right)^2 = 2\int_{0}^{\infty}\ud t\,\frac{w(t^2)}{t}\,\left(J_{n}(2t)\right)^2 = \frac{1}{n^2}\,. 
\end{equation}
Let us examine the following integral formula for Bessel functions 
\cite{magnus66}, 
\begin{equation}
\label{intbessel2}
\int_0^{\infty}\frac{\ud t}{t} \, \left(J_{n}(2 t)\right)^2= \frac{1}{2n}\, .  
\end{equation}
This leads us to replace the SG-CS of \eqref{SGmoya} by the modified
\begin{equation}
\label{SGm}
|\alpha\rg_{\mathrm{SGm}}= \sum_{n=0}^{\infty}\alpha^n\,h^{\mathrm{SGm}}_n(\vert\alpha\vert^2) \,|n\rg\,, \quad h^{\mathrm{SGm}}_n(u) = \sqrt{\frac{n+1}{\mathcal{N}(u)}}\,\frac{1}{u^{\frac{n+1}{2}}}\,J_{n+1}(2\sqrt{u})\, ,
\end{equation}
with 
\begin{equation}
\label{normSGm} 
\mathcal{N}(u) = \frac{1}{u}\sum_{n=1}^{\infty}  n\,\left(J_{n}(2\sqrt{u})\right)^2\,.
\end{equation}
Then the formula \eqref{intbessel2} allows us to prove that the resolution of the identity is fullfilled by these $|\alpha\rg_{\mathrm{SGm}}$  with $w(u)= \mathcal{N}(u) $. 
%Let us examine the following integral formula for Bessel functions 
%\cite{magnus66}, 
%\begin{equation}
%\label{intbessel2}
%\int_0^{\infty}\ud x \,x\,e^{-cx^2}\, J_{\nu}(a x)\,J_{\nu}(b x)=  \frac{1}{2c}\,e^{-\frac{a^2 +b^2}{4c}}\,I_{\nu}\left(\frac{ab}{2c}\right)\, , 
%\end{equation}
%where $I_{\nu}$ is a modified Bessel function. Then let us compare  the above integral with $\nu = n+1$ and $a=b=2$ with  \eqref{mom}  applied to the present case, 
%\begin{equation}
%\label{SGweight1}
% \int_{0}^{\infty}\ud x\,w(x)\,x^n\vert h_n(x)\vert^2=\frac{1}{(n+1)^2}\int_0^{\infty}\ud x \,\frac{w(x)}{x}\left(J_{n+1}(2x)\right)^2\, . 
%\end{equation}
%So, if we put $w(x) = x^2 \,e^{-cx^2}$, we find from \eqref{intbessel2},
%\begin{equation}
%\label{SGweight2}
%\frac{1}{(n+1)^2}\int_0^{\infty}\ud x \,x\,e^{-cx^2}\left(J_{n+1}(2x)\right)^2= \frac{1}{2c(n+1)^2}\,e^{-\frac{2}{c}}\,I_{n+1}\left(\frac{2}{c}\right)\,. 
%\end{equation}
%So, it is very doubtful to find $c$ such that the r.h.s. of this equation be equal to $1$!
%\subsection{Laguerre CS}
More details, particularly those concerning statistical aspects, are given in \cite{cufagano18}.

\section{CS from symmetric deformed binomial distributions (DFB)}
\label{DFBCS}

In \cite{bercugaro13} (see also the related works \cite{cugaro10,cugaro11,bercugaro12}) was presented  the following generalization of the binomial distribution,
\begin{equation}
\label{genbinsym}
\mathfrak{p}_k^{(n)}(\xi)=\dfrac{x_n!}{x_{n-k}! x_k!} q_k(\xi) q_{n-k}(1-\xi)\, ,
\end{equation}
where the $\{x_n\}$'s form a non-negative sequence and the $q_k(\xi)$ are polynomials of  degree $k$, while $\xi$ is a running parameter on the interval $[0,1]$. The $\mathfrak{p}_k^{(n)}(\xi)$ are constrained by:
\begin{itemize}
\item[(a)]the normalization
\begin{equation}
\label{eq:polynorm}
\forall n \in \mathbb{N}, \quad \forall \xi \in [0,1], \quad \sum_{k=0}^n \mathfrak{p}_k^{(n)}(\xi)=1,
\end{equation}
\item[(b)]the non-negativeness condition (requested by statistical interpretation)
\begin{equation}
\label{poscondsym}
\forall n,k \in \mathbb{N}, \quad \forall \xi \in [0,1], \quad \mathfrak{p}_k^{(n)}(\xi) \ge 0.
\end{equation}
\end{itemize}
These conditions imply that $q_0(\xi)=\pm 1$. With the choice  $q_0(\xi)=1$ one easily proves that  the non-negativeness condition \eqref{poscondsym} is equivalent to the non-negativeness of the polynomials $q_n$ on the interval $[0,1]$.
Hence the quantity $\mathfrak{p}_k^{(n)}(\xi)$  can be interpreted as the probability of having $k$ wins and $n-k$ losses in a sequence of \emph{correlated} $n$ trials.  Besides, as we recover the  invariance under  $k \to n-k$ and $\xi \to 1-\xi$ of the binomial distribution, no bias (in the case $\xi = 1/2$) can exist favoring either win or loss.
The polynomials $q_n(\xi)$ are viewed here as \emph{deformations} of $\xi^n$. 
We now suppose that the generating function for the polynomials $q_n$, defined as
\begin{equation}
\label{genfctgen}
F(\xi; t) := \sum_{n=0}^\infty \dfrac{q_n(\xi)}{x_n!} t^n\, , 
\end{equation}
can be expressed as
\begin{equation}
\label{ }
F(\xi; t) = e^{\sum_{n=1}^\infty a_n t^n} \quad \mbox{with} \quad a_1=1\,,\, a_n=a_n(\xi) \ge 0\,, \, \sum_{n=1}^\infty a_n < \infty\, . 
\end{equation}
It is proved in \cite{bercugaro13} that conditions of normalization (a) and non-negativeness (b) on $\mathfrak{p}_k^{(n)}(\xi)$ are satisfied. 
We now define
\begin{equation}
\label{deffandb}
f_n = \int_0^\infty q_n(\xi) \, e^{-\xi} \,\ud\xi \quad \textrm{and} \quad b_{m,n}=\int_0^1 q_m(\xi) \,q_n(1-\xi)\, \ud\xi \,.
\end{equation}
The $f_n$ and $b_{m,n}$ are  deformations of the usual factorial and beta function, respectively deduced from their usual integral definitions through the substitution $\xi^n \mapsto q_n(\xi)$.  The following properties are proven in \cite{bercugaro13}:
\begin{equation}
\label{fbproperties}
\begin{split}
q_n(\xi)&\geq 0 \ \forall \xi \in \mathbb{R}^+\, , \quad x_n! \le f_n\, , \\    \sum_{n=0}^\infty &\dfrac{q_n(\xi)}{f_n} < \infty\ \forall \xi \in \mathbb{R}^+\, ,  \quad \mbox{and}\quad b_{m,n} \ge \dfrac{x_m! x_n!}{(m+n+1)!} \,.
\end{split}
\end{equation}
Then let us introduce the function $\mathcal{N}(z)$ defined on $\mathbb{C}$ as
\begin{equation}
\forall z \in \mathbb{C}\quad \mathcal{N}(z) = \sum_{n=0}^\infty \dfrac{q_n(z)}{f_n} \,.
\end{equation}
This definition makes sense since from Eq. \eqref{fbproperties}
\begin{equation}
\sum_{n=0}^\infty \left| \dfrac{q_n(z)}{f_n} \right| \le \sum_{n=0}^\infty \dfrac{q_n(|z|)}{f_n} < \infty .
\end{equation}
The above material allows us to present two new  generalisations of standard and spin coherent states.

\subsubsection*{ DFB coherent states on the complex plane}
They are  defined in the Fock space  as
\begin{equation}
| \alpha \rangle_{\mathrm{dfb}} = \dfrac{1}{\sqrt{\mathcal{N}(|\alpha|^2)}} \sum_{n=0}^\infty \dfrac{1}{\sqrt{f_n}} \sqrt{q_n(|\alpha|^2)}\, e^{\ii \, n \arg(\alpha)} |n \rangle \,.
\end{equation}
These states verify the following resolution of the unity:
\begin{equation}
\int_{\mathbb{C}} \dfrac{\ud^2\alpha}{\pi} e^{- |\alpha|^2} \mathcal{N}(|\alpha|^2) \, |\alpha \rangle_{\mathrm{dfb}}{}_{\mathrm{dfb}} \langle \alpha | = I \, .
\end{equation}
They are a natural generalization of the standard coherent states that correspond to the special polynomials $q_n(\xi)= \xi^n$. The latter are associated to the generating function $F(t)=e^t $ that gives the usual binomial distribution.

\subsubsection*{DFB spin coherent states}

These states can be considered as generalizing the  spin coherent states \eqref{zetaspincs}.   
\begin{equation}
\label{genspinCS}
| \alpha;n_j\rg_{\mathrm{dfb}}= \dfrac{1}{\sqrt{\mathcal{N}(\vert\alpha\vert^2)}} \sum_{n=0}^{n_j}\sqrt{\frac{q_{n} \left( \frac{1}{1+\vert \alpha\vert^2}\right) q_{n_j-n} \left( \frac{\vert\alpha\vert^2}{1+\vert \alpha\vert^2} \right) }{b_{n,n_j-n}}} e^{\ii \arg(\alpha)}|n\rg\, , 
\end{equation}
where the $b_{m,n}$ are defined in Eq. \eqref{deffandb} and $\mathcal{N}(u)$ is given by
\begin{equation}
\mathcal{N}(u) = \sum_{n=0}^{n_j}\frac{q_{n} \left( \frac{1}{1+u}\right) q_{n_j-n} \left( \frac{u}{1+u} \right) }{b_{n,n_j-n}}\,.
\end{equation}
The family of states (\ref{genspinCS}) resolves the unity:
\begin{equation}
\label{resunitj}
%\frac{1}{4\pi}\int_{S^2} \sin\theta\, d\theta\, d\phi \, \varpi (\theta) \, | \theta,\phi\rg\lg  \theta,\phi| = 1_d\, . 
 \int_{\C}\ud^2\alpha\,\mathfrak{w}\left(\alpha\right)\, | \alpha;n_j\rg_{\mathrm{dfb}}{}_{\mathrm{dfb}}\lg\alpha;n_j| = I\, , \quad \mathfrak{w}\left(\alpha\right)= \frac{\mathcal{N}\left(\vert \alpha\vert^2\right) }{\pi\left(1+\vert \alpha\vert^2\right)^2}\,. 
\end{equation}

\section{Photon-counting, basic statistical aspects}
\label{stat}
In this section, we mainly follow  the inspiring chapter 5 of Ref. \cite{fox06} (see also  the seminal papers \cite{mandel58,mandelwolf70,klyshko96} on the topic, the renowned \cite{loudon00}, the pedagogical \cite{koczyk96}, and the more recent \cite{gerry04,bachor04,eimigpol11}).  In quantum optics one views a beam of light as a stream of discrete energy packets named ``photons'' rather than a classical wave. With a photon counter the average count rate is determined by the intensity of the light beam, but the actual count rate fluctuates from measurement to measurement. Whence, one easily understands that two statistics are in competition here, on one hand the statistical nature of the photodetection process, and on the other hand, the intrinsic photon statistics of the light beam, e.g., the average $\bar n(\alpha)$ for a CS $|\alpha\rg$. 
Photo-counting detectors are specified by their quantum efficiency $\eta$, which is defined as the ratio of the number of photocounts to the number of incident photons. For a perfectly coherent monochromatic beam of angular frequency $\omega$, constant intensity $I$ and area $A$, and for a counting time $T$, 
\begin{equation}
\label{etaef}
\eta = \frac{N(T)}{\Phi T}\, ,  
\end{equation} 
where the photon flux is $\Phi= \dfrac{IA}{\hbar \omega}\equiv \dfrac{P}{\hbar \omega}$, $P$ being the power. Thus the corresponding count rate is $\mathcal{R} = \dfrac{\eta P}{\hbar \omega}$ counts s$^{-1}$. Due to a ``dead time" of $\sim 1\,\mu$s for the detector reaction, the count rate cannot be larger than $\sim 10^{6}$ counts  s$^{-1}$, and due to weak values $\eta \sim 10\%$ for standard detectors, photon counters are only useful for analysing properties of very faint beams with optical powers of $\sim 10^{-12}$W or less. The detection of light beams with higher  powers requires other methods. 

Although the average photon flux can have a well-defined value, the photon number on short time-scales fluctuates due to the discrete nature of the photons. These fluctuations are described by the photon statistics of the light. 

One proves that the photon statistics for a coherent light wave with constant intensity (e.g., a light beam described by the electric field $\mathcal{E}(x,t)= \mathcal{E}_0 \sin(kx-\omega t + \phi)$  with constant angular frequency $\omega$, phase $\phi$, and  intensity $\mathcal{E}_0$)  is encoded by the Poisson distribution 
\begin{equation}
\label{photstatpoisson}
n\mapsto \sfP_n(\bar n)= e^{-\bar n}\, \frac{(\bar n)^n}{n!}\, ,
\end{equation}
This randomness of the count rate of a photon-counting system detecting individual photons from a light beam with constant intensity  originates from chopping the continuous beam into discrete energy packets with an equal probability of finding the energy packet within any given time subinterval.

Let us introduce the variance as the quantity
\begin{equation*}
\mathrm{Var}_n(\bar n)\equiv (\Delta n)^2= \sum_{n=0}^{\infty} (n-\bar n)^2 \sfP_n(\bar n)\, , 
\end{equation*}
Thus, for a Poissonian coherent beam, $\Delta n= \sqrt{\bar n}$. There results that three different types of 
photon statistics can occur: Poissonian, super-Poissonian, and sub-Poissonian. The two first ones are consistent as well with the classical theory of light, whereas sub-Poissonian statistics is not and constitutes direct confirmation of the photon nature of light. More precisely, 
\begin{itemize} 
\item[(i)] if the  Poissonian statistics holds, e.g., for a perfectly coherent light beam with constant optical power $P$, we have
 \begin{equation}
 \label{poissdeln}
\Delta n= \sqrt{\bar n}\, , 
\end{equation}
\item[(ii)] if the super-Poissonian statistics, e.g., classical light beams with time-varying light intensities, like thermal light from a black-body source, or like partially coherent light from a discharge lamp, we have
\begin{equation}
\label{superpoiss}
\Delta n > \sqrt{\bar n}\, , 
\end{equation}
\item[(iii)]  finally, the sub-Poissonian statistics is featured by a narrower distribution than the Poissonian case
\begin{equation}
\label{subpoiss}
\Delta n < \sqrt{\bar n}\,.
\end{equation}
This light  is ``quieter'' than the perfectly coherent light. Since a perfectly coherent beam is the most stable form of light that can be envisaged in classical optics,  sub-Poissonian light has no classical counterpart. 
\end{itemize} 
In this context popular useful parameters are introduced to account for CS statistical properties, e.g. the Mandel parameter $Q= (\Delta n)^2/\bar n -1$, where $(\Delta n)^2= \overline{n^2} -\bar n^2$,  which is $<0$ (resp. $>0$, $=0$) for sub-Poissonian (resp. super-Poissonian, Poissonian), the  parameter $Q/\bar n +1$ which is $>1$  for  ``bunching'' CS and $<1$ for ``anti-bunching'' CS, etc..

The aim of the  quantum theory of photodetection  is to relate the photocount statistics observed in a particular experiment to those of the incoming photons, more precisely  the average photocount number $\bar N$ to the mean photon number $\bar n$ incident on the detector in a same time interval.
The quantum efficiency $\eta$ of the detector, defined as $\eta=\bar N/\bar n$
 is the critical parameter that determines the relationship between the photoelectron and photon statistics. 
  Indeed, consider the relation between variances $(\Delta N)^2 = \eta^2 \,(\Delta n)^2 + \eta \,(1-\eta)\,\bar n $.
\begin{itemize}
  \item If $\eta = 1$, we have $\Delta N = \Delta n$: the photocount fluctuations faithfully reproduce the fluctuations of the incident photon stream.
\item If the incident light has Poissonian statistics  $\Delta n = \sqrt{\bar n}$ then $(\Delta N)^2 =  \eta \,\bar n$  for all values of $\eta$: photocount is  Poisson.
  \item If $\eta\ll 1$, the photocount fluctuations tend to the Poissonian result with $(\Delta N)^2 =  \eta \,\bar n = \bar N$  irrespective of the underlying photon statistics.
\end{itemize}
Observing sub-Poissonian statistics in the laboratory is a delicate matter since it depends on the availability of single-photon detectors with high quantum efficiencies.

\section{AN CS quantization}
\label{ANCSQ}
\subsection{The quantization map and its complementary}
If the resolution of the identity \eqref{resid} is valid for a given family of AN CS determined by the sequence of functions $\boldsymbol{\mathsf{h}}:=\left(h_n(u)\right)$, it makes the quantisation of functions (or distributions) $f(\alpha)$ possible along the linear map
\begin{equation}
\label{ANCSquant}
f(\alpha) \mapsto A^{\bsh}_f= \int_{\vert \alpha\vert < R}\frac{\ud^2\alpha}{\pi} \,w(\vert \alpha\vert^2)\, f(\alpha)\,|\alpha\rg\lg\alpha|\,, 
\end{equation}
together with its complementary map,  likely to provide a ``semi-classical'' optical phase space portrait, or \textit{lower symbol},  of $A^{\bsh}_f$ through the map \eqref{mapaln},
\begin{equation}
\label{phspport}
\lg\alpha|A^{\bsh}_f|\alpha\rg = \int_{\vert \beta\vert < R}\frac{\ud^2\beta}{\pi} \,w(\vert \beta\vert^2)\, f(\beta)\,\vert\lg\alpha|\beta\rg\vert^2\equiv \widecheck{f^{\bsh}}(\alpha)\,. 
\end{equation}
Since for fixed $\alpha$ the map $\beta \mapsto w(\vert \beta\vert^2)\, \vert\lg\alpha|\beta\rg\vert^2$ is a probability distribution on the centered disk $\mathcal{D}_{R}$ of radius $R$, the map $f(\alpha) \mapsto \widecheck{f^{\bsh}}(\alpha)$ is a local, generally regularising, averaging, of the original $f$. 

The quantization map \eqref{ANCSquant} can be extended to cases  comprising geometric constraints in the optical phase portrait through the map \eqref{mapaln}, and encoded by distributions like Dirac or Heaviside functions. 

\subsection{AN CS quantization of simple functions}
When applied to the  simplest functions $\alpha$ and $\bar\alpha$ weighted by a positive $\mathfrak{n}\left(\vert\alpha\vert^2\right)$,   the quantization map \eqref{ANCSquant} yields lowering and raising operators
\begin{align}
\label{lowering}
   \alpha & \mapsto   a^{\bsh}= \int_{\vert \alpha\vert < R}\frac{\ud^2\alpha}{\pi} \,\tilde w(\vert \alpha\vert^2)\, \alpha\,|\alpha\rg\lg\alpha|= \sum_{n=1}^{\infty}a^{\bsh}_{n-1 n}|n-1\rg\lg n|\, , \\
 \label{raising}   \bar \alpha&\mapsto \left(a^{\bsh}\right)^{\dag} =  \sum_{n=0}^{\infty}\overline{a^{\bsh}_{n n+1}}|n+1\rg\lg n|\, , 
\end{align}
where $\tilde w(u):= \mathfrak{n}(u)w(u)$.
Their matrix elements are given by the integrals
\begin{equation}
\label{ann}
a^{\bsh}_{n-1 n}:= \int_0^{R^2}\ud u\, \tilde w(u)\,u^n\, h_{n-1}(u)\, \overline{h_n(u)}\, , 
\end{equation}
and $a^{\bsh}|0\rg = 0$. 

The lower symbol of $a^{\bsh}$ and its adjoint read respectively:
\begin{equation}
\label{checkaa}
\widecheck{a^{\bsh}} (\alpha)=\lg \alpha| a^{\bsh} |\alpha\rg = \alpha \,\tau\left(\vert \alpha\vert^2\right)\, , \quad \widecheck {\adgh}(\alpha)
= \overline{\widecheck{a^{\bsh}} (\alpha)}\, , 
\end{equation}
in which the ``weighting''  factor is given by $\tau(u)= \sum_{n\geq 0} a^{\bsh}_{n n+1}\,u^n\, \overline{h_{n}(u)}\, h_{n+1}(u)$. 

 In the above, as it was mentioned in Section \ref{genset} and, as  it occurred in the spin case, the involved sums can be finite, and a finite number of matrix elements \eqref{ann} are not zero. 
As a generalization of the number operator we get in the present case
\begin{equation}
\label{comrule}
a^{\bsh}\adgh= \sfX^{\bsh}_{\hN+I}\, , \quad \adgh a= \sfX^{\bsh}_{\hat N}\,, \quad  [a^{\bsh},\adgh]= \sfX^{\bsh}_{\hat N+I}-\sfX^{\bsh}_{\hat N}\, , 
\end{equation}
with the notations
\begin{equation}
\label{xn}
\sfX^{\bsh}_n = \vert a^{\bsh}_{n-1 n}\vert^2\, ,\quad  \sfX^{\bsh}_0= 0\, ,\quad \sfX^{\bsh}_{\hN} |n\rg= \sfX^{\bsh}_n|n\rg\, , \quad \sfX^{\bsh}_{\hN +I}|n\rg = \sfX^{\bsh}_{n+1}|n\rg\,.
\end{equation}
When all the $h_n$'s are real, the diagonal elements in  \eqref{comrule} are given by the product of integrals
\begin{equation}
\label{comruleR}
\begin{split}
\sfX^{\bsh}_{n+1} -\sfX^{\bsh}_n&= \left[\int_0^{R^2}\ud u\, \tilde w(u)\,u^n\,h_n(u)\,\left(uh_{n+1}(u)-h_{n-1}(u)\right)\right]\\&\times\left[\int_0^{R^2}\ud u\, \tilde w(u)\,u^n\,h_n(u)\,\left(uh_{n+1}(u)+h_{n-1}(u)\right)\right]\,.
\end{split} 
\end{equation}
 The quantum version of $u=\vert \alpha\vert^2$ and its lower symbol read as:
\begin{equation}
\label{Ahu}
\begin{split}
A^{\bsh}_{u}&= \sum_n \lg u \rg_n |n\rg\lg n|\, , \quad \lg u \rg_n:= \int_0^{R^2}\ud u\, \tilde w(u)\,u^{n+1}\,h_n(u)\\
\lg \alpha| A^{\bsh}_{u} |\alpha\rg &= \left\lg \lg u \rg_n\right\rg_{\alpha}(u):= \sum_n  \lg u \rg_n \, u^n\,\vert h_n(u)\vert^2= \sum_n  \lg u \rg_n \,\sfP^{\bsh}_n\, . 
\end{split}
\end{equation}
We notice here an interesting   duality between classical ($\lg\cdot\rg_n$) and quantum ($\left\lg \cdot\right\rg_{\alpha}$) statistical averages.

\subsection{AN CS as $a$-eigenstates}
One crucial property of the Glauber-Sudarshan CS is that they are eigenstates of the lowering operator $a$.  Imposing this property to AN-CS leads to a supplementary condition on  the functions $h_n$. 
\begin{equation}
\label{aa=aa}
a^{\bsh}|\alpha\rg = \alpha|\alpha\rg \Rightarrow h_n(u)=h_{n+1}(u)\,\int_0^{R^2}\ud t \,\tilde w(t) \,t^{n+1}\, h_n(t)\,\overline{h_{n+1}(t)}\, . 
\end{equation}
Let us examine the particular case of non-linear CS of the deformed Poissonian type \eqref{DPNLCS}. In this case, $\sfX_n=x_n$, and whence the construction formula,
\begin{equation}
\label{constrxn}
|\alpha\rg = \frac{\mathcal{N}(\alpha {a^{\bsh}}^{\dag})}{\sqrt{\mathcal{N}(\vert \alpha \vert^2)}}|0\rg\, .
\end{equation}
  Morover \eqref{aa=aa} imposes that the sequence $x_n!$  derives from  the following moment problem.
\begin{equation}
\label{momxn}
x_n! = \int_0^{R^2}\ud u \, \frac{w(u)}{\mathcal{N}(u)}\,u^n\, . 
\end{equation}
Now, instead of starting from a known sequence $(x_n)$, one can reverse the game by choosing a suitable  function $f(u)= \dfrac{w(u)}{\mathcal{N}(u)}$ to calculate the corresponding $x_n!$ (from which we deduce the $x_n$'s), the resulting generalized exponential $\mathcal{N}(u)$ (and checking the finiteness of the convergence radius), and eventually the weight function $w(u)=f(u)\,\mathcal{N}(u)$. There are an infinity of ``manufactured" products in this non-linear CS factory!
 
 \subsection{AN-CS from displacement operator}
 One can attempt to build (other?) AN-CS by following the standard procedure involving the unitary ``displacement'' operator built from $a^{\bsh}$ and ${a^{\bsh}}^{\dag}$ and acting on the vacuum.
 \begin{equation}
\label{CDdispv}
|\breve\alpha\rg_{\mathrm{disp}}:= D_{\bsh}(\breve\alpha) \, |0\rg = \sum_{n=0}^{\infty}\breve \alpha^n\,h_n^{\mathrm{disp}}(\vert\breve \alpha\vert^2)\,|n\rg\, , \quad D_{\bsh}(\breve\alpha):= e^{\breve\alpha {a^{\bsh}}^{\dag} -\overline{\breve\alpha} a^{\bsh}}\, , 
\end{equation}
where the notation $\breve\alpha$ is used to make the distinction from the original $\alpha$.  Of course, ${D_{\bsh}}^{\dag}(\breve\alpha)= {D_{\bsh}}^{-1}(\breve\alpha)$ is not equal in general to $D_{\bsh}(-\breve\alpha)$. Besides the two examples \eqref{SU2disp} and \eqref{SU11disp} encountered in the SU$(2)$ and SU$(1,1)$ CS constructions, for which the respective weights $\mathfrak{n}(u)$ can be given explicitly, another  recent interesting example  is given in \cite{rodriguezlara17}. 

  So an appealing program is to establish the relation between the original $h_n$'s and these (new?) $h_n^{\mathrm{disp}}$'s, through a suitable choice of the weight $\mathfrak{n}(u)$, actually a big challenge in the general case!  More interesting yet is the fact that  these new CS's might be experimentally produced 
in the Glauber's way \eqref{vactot}, once we accept that the $a^{\bsh}$ and ${a^{\bsh}}^{\dag} $ appearing in the quantum version \eqref{FourexpA} of the classical e.m. field are yielded by a CS quantization different from the historical Dirac (canonical) one \cite{dirac27a}. Hence one introduces a kind of duality between two families of coherent states, the first one used in the quantization procedure $f(\alpha) \mapsto A^{\bsh}_f$, producing the operators $\mathfrak{n}(u)\alpha \mapsto a^{\bsh}$ and $\mathfrak{n}(u)\bar\alpha \mapsto {a^{\bsh}}^{\dag}$, and so the unitary displacement $D^{\bsh}(\breve\alpha):= e^{\breve\alpha {a^{\bsh}}^{\dag} -\overline{\breve\alpha} a^{\bsh}}$, while the other one uses this $D_{\bsh}(\breve\alpha)$ to build potentially experimental CS yielded in the Glauber's way.

\section{Conclusion}
\label{conclu}
We have presented in this paper a  unifying approach to build coherent states in a wide sense that are potentially relevant to quantum optics. Of course, for most of them, their experimental observation or production comes close to being impossible with the current experimental 
physics. Nevertheless, when one considers the way quantum optics has emerged from the golden twenties of quantum mechanics,
nothing prevents us to enlarge the Dirac quantization of the classical e.m. field in order to include all these deformations (non-linear or others) by adopting the consistent method exposed in the previous section.

\subsection*{Acknowledgments} This research is supported in part by the Ministerio de Econom\'ia y Competitividad of Spain  under grant  MTM2014-57129-C2-1-P and the Junta de Castilla y Le\'on (grant VA137G18). The author is also indebted to the University of Valladolid. He thanks M. del Olmo (UVA) for helpful discussions about this review. He addresses special thanks to Y. Hassoumi (Rabat University) and to the Organizers of the Workshop QIQE'2018 in Al-Hoceima, Morocco for valuable comments and questions which allowed to improve significantly the content of this review.


\begin{thebibliography}{99}


\bibitem{kindaoud01} A.~H. El Kinani and M. Daoud, Generalized intelligent states for an arbitrary quantum system, 
\textit{J. Phys. A: Math. Gen.} \textbf{34} (2001) 5373-5387.  

\bibitem{hach_etal16} E.~E. Hach III, P.~M. Alsing, and C.~C. Gerry, Violations of a Bell inequality for entangled SU$(1,1)$ coherent states based on dichotomic observables, \textit{Phys. Rev. A} \textbf{93} (2016) 042104-1-8.

 \bibitem{cruzgress17} S. Cruz y Cruz and Z. Gress, Group approach to the paraxial propagation of Hermite-Gaussian modes in a parabolic medium, \textit{Ann.  Phys. (NY)} \textbf{383} (2017) 257-277.

\bibitem{hohumaz18} S.~E. Hoffmann, V. Hussin, I. Marquette, and Yao-Zhong Zhang, Non-classical behaviour of coherent states  for systems constructed using exceptional  orthogonal polynomials, \textit{J. Phys. A: Math. Theor.}
\textbf{51} (2018) 085202-1-16.

\bibitem{gohoza12} K. G\'orska, A. Horzela, and F.~H. Szafraniec,
Coherence, Squeezing and Entanglement: An Example of Peaceful Coexistence,  in \cite{anbaga18}
 89-117.

\bibitem{hach_etal18} E.~E. Hach, R. Birrittella, P.~M. Alsing, and C.~C. Gerry, SU(1,1) parity and strong violations of a Bell inequality by entangled Barut-Girardello coherent states, \textit{J. Opt. Soc. Am. B} \textbf{35} (2018)  2433-2442. 

\bibitem{glauber63-1} R.~J. Glauber,
          Photons correlations,  \textit{Phys. Rev. Lett.} \textbf{10} 
            (1963) 84-86. 
            
 \bibitem{gasza11} J.-P. Gazeau1 and F.~H. Szafraniec, Holomorphic Hermite polynomials and a non-commutative plane,  \textit{J. Phys. A: Math. Theor.} \textbf{44} (2011) 495201-1-13. 

\bibitem{magnus66}    W. Magnus, F. Oberhettinger, and R.~P.  Soni.
\newblock {\em Formulas and Theorems for
the Special Functions of Mathematical Physics}.
 \newblock Springer-Verlag,  Berlin, Heidelberg and New York, 1966.
 
 \bibitem{schwinger53} J. Schwinger, The Theory of Quantized Fields. III,
\textit{Phys. Rev.} \textbf{91} (1953) 728-740.


\bibitem{glauber63-2} R.~J. Glauber, 
          The quantum theory of optical coherence,  \textit{Phys. Rev.} \textbf{130} 
            (\textbf{1963}) 2529-2539. 

\bibitem{glauber63-3} R.~J. Glauber, 
          Coherent and incoherent states of radiation field,  \textit{Phys. Rev.} \textbf{131} 
            (1963) 2766-2788.

\bibitem{sudarshan63} E.~C.~G. Sudarshan,  Equivalence of semiclassical and quantum 
		mechanical   descriptions  of statistical light beams,   
		\textit{Phys. Rev. Lett.} \textbf{10} (1963) 277-279.
		
\bibitem{mandel_wolf65} L. Mandel and E. Wolf, Coherence Properties of Optical Fields,
\textit{Rev. Mod. Phys.} \textbf{37}, (1965) 231-287.

\bibitem{cahglau69} K.~E. Cahill and R. J. Glauber, Ordered Expansions in Boson Amplitude Operators,
\textit{Phys. Rev.} \textbf{177} (1969) 1857-1881. 

\bibitem{agawo70} B. S. Agarwal and E. Wolf, Calculus for Functions of Noncommuting Operators and General Phase-Space Methods in Quantum Mechanics, \textit{Phys. Rev. D} \textbf{2} (1970) 2161-2186 (I), 2187-2205 (II), 2206-2225 (III).

\bibitem{schrodinger26} Schr\"odinger, E. : Der stetige \"Ubergang von der
Mikro- zur Makromechanik, \textit{Naturwiss.} 14 (\textbf{1926}) 664.

\bibitem{klauder60}  J.~R. Klauder,  The Action Option and the Feynman Quantization of Spinor Fields in Terms of Ordinary c-Numbers, \textit{ Annals of Physics} \textbf{11} (1960) 123.

\bibitem{klauder63-1} J.~R. Klauder, Continuous-Representation Theory I.
Postulates of continuous-representation theory, \textit{J. Math. Phys.}
\textbf{4} (1963) 1055-1058.

\bibitem{klauder63-2} J.~R. Klauder,  Continuous-Representation Theory II.
Generalized relation between quantum and classical dynamics, \textit{J.
Math. Phys.} \textbf{4} (1963) 1058-1073.

\bibitem{KlauSkag85} J.~R. Klauder and B.~S Skagerstam, editors.
 {\em  Coherent states. 
Applications in physics and mathematical physics}. 
World Scientific Publishing Co., Singapore, 1985. 

\bibitem{perel72} A.~M. Perelomov,  Coherent States for Arbitrary Lie Group, \textit{Commun. math. Phys.} \textbf{26}, (1972) 222-236.  

\bibitem{perel86} A.~M. Perelomov, \textit{Generalized Coherent States and Their Applications} (Springer, Berlin, 1986).

\bibitem{zhangfenggil90}  W-M. Zhang, D.~H. Feng, and R. Gilmore,  Coherent states: Theory and some  applications,   \textit{Rev. Mod. Phys.} \textbf{26} (1990) 867-927.

\bibitem{FengKlau94}  D.~H. Feng,  J.~R. Klauder, and  
M. Strayer, editors.  {\em Coherent States: Past, Present and Future}. 
Proceedings of the 1993 Oak Ridge Conference.
World Scientific, Singapore, 1994.

\bibitem{aagbook14} S.~T. Ali,  J.-P Antoine, and J.-P. Gazeau, \textit{Coherent States, Wavelets and their Generalizations\/} (2000),   2d edition, Theoretical and Mathematical Physics, Springer, New York (2014). 	

\bibitem{dodonov02} V.~V. Dodonov, `Nonclassical' states in quantum optics: a `squeezed' review of the first 75 years, 
   \textit{J. Opt. B: Quantum Semiclass. Opt.} \textbf{4} (2002) R1.

\bibitem{dodoman03} V.~V. Dodonov and V.~I. Man'ko, editors. 
{\em  Theory of Nonclassical States of Light}.
 Taylor \& Francis, London , New York, 2003. 

\bibitem{vourdas06} A. Vourdas,  Analytic representations in quantum mechanics, {\it J. Phys.
A}  \textbf{39} (2006) R65.

\bibitem{gazeaubook09} J.-P. Gazeau,  
\textit{Coherent States in Quantum Physics},  Wiley-VCH, Berlin, 2009.


\bibitem{alanbaga12} S.T. Ali, J.P. Antoine, F. Bagarello, and J.~P. Gazeau, \textit{Special issue on coherent states: mathematical and physical aspects},  \textit{J. Phys. A: Math. Theor.} \textbf{45} (2012).

\bibitem{anbaga18} J.-P. Antoine, F. Bagarello, and J.~P. Gazeau, \textit{Coherent States and their applications: A contemporary panorama},  Proceedings of the CIRM workshop November 13-18, 2016,  Springer Proceedings in Physics (SPPHY) \textbf{205} (2018).

\bibitem{cotgagor10} N. Cotfas, J.-P. Gazeau, K. G\'orska, Complex and real Hermite polynomials and related quantizations,\textit{J. Phys. A: Math. Theor.} \textbf{43} (2010) 305304-1-14.

\bibitem{albaga12} S.~T. Ali, F. Bagarello, J.-P.  Gazeau,  Quantizations from reproducing kernel spaces,  \textit{Ann. Phys. (NY)} \textbf{332} (2012) 127-142. 

\bibitem{gazolmo13} J.-P. Gazeau and M.~A. del Olmo, Pisot $q$-coherent states quantization of the harmonic oscillator, \textit{Ann. Phys. (NY)} \textbf{330} (2013) 220-245. 

\bibitem{solkac}  A. De Sole and V.  Kac, On integral representations of $q$-gamma and $q$-beta functions, \emph{Rend. Mat. Acc. Lincei} {\bf 9} (2005) 11-29;    arXiv: math.QA/0302032.

\bibitem{bafregaha12} M. El Baz, R. Fresneda, J.-P. Gazeau and Y. Hassouni,  Coherent state quantization of paragrassmann algebras, \textit{ J. Phys. A: Math. Theor.} \textbf{43} (2010) 385202-1-15; Corrigendum
\textit{ J. Phys. A: Math. Theor. }  \textbf{45} (2012) 079501-1-2.


\bibitem{fox06} M. Fox, \textit{Quantum Optics: An Introduction}, Oxford University Press, New York, 2006.

\bibitem{algahel08} S.~T. Ali, J.-P. Gazeau, and B. Heller,  Coherent states and Bayesian duality, \textit{ J. Phys. A: Math. Theor.} \textbf{41} (2008) 365302-1-22. 

\bibitem{gahulare07} J.-P. Gazeau, E. Huguet, M. Lachi\`eze-Rey, and J. Renaud, Fuzzy spheres from inequivalent coherent states quantizations, \textit{J. Phys. A: Math. Theor.} (2007) \textbf{40} 10225-10249.

\bibitem{jordan35} P. Jordan,(1935)Der Zusammenhang der symmetrischen und linearen Gruppen und das Mehrk\"{o}rperproblem", \textit{Zeitschrift f\"{u}r Physik} \textbf{94} (1935) 531-535. 

\bibitem{holsprim40} T. Holstein and H. Primakoff, \textit{Phys. Rev.} \textbf{58} (1940) 1098-1113.

\bibitem{schwinger52} J. Schwinger, On Angular Momentum, Unpublished Report, Harvard University, Nuclear Development Associates, Inc., United States Department of Energy (through predecessor agency the Atomic Energy Commission), Report Number NYO-3071 (January 26, 1952).

\bibitem{gazolmo18} J.-P. Gazeau and M. del Olmo, Covariant integral  quantization of the unit disk, \textit{in preparation}.

\bibitem{aharonov73} Y. Aharonov, E.~C. Lerner, H.~W. Huang, and J.~M. Knight, Oscillator phase states, thermal equilibrium and group representations, \textit{J. Math. Phys.} \textbf{14} (2011) 746-755.

\bibitem{bargir71} A.~O. Barut and L. Girardello, New ``Coherent'' States Associated with Non-Compact Groups, \textit{Commun. Math. Phys.} \textbf{21} (1971) 41-55.

\bibitem{angaklamopen01} J.-P. Antoine, J.-P. Gazeau, J.~R. Klauder,  P. Monceau, and K.~A. Penson, \textit{J. Math. Phys.} \textbf{42} (2001) 2349-2387.

\bibitem{sussglog64} L. Susskind and J. Glogower, Quantum mechanical phase and time operator,
 \textit{Phys. Phys. Fiz. 1} \textbf{1}  (1964) 49-61.
 
 \bibitem{moyasoto11} H.~M. Moya-Cessa and F. Soto-Eguibar, \textit{Introduction to Quantum Optics}, Rinton Press, Paramus, 2011. 
 
 \bibitem{evaldo-6-18} Evaldo Curado, \textit{private communication}.

\bibitem{cufagano18} E.~M.~F. Curado, S. Faci, J.-P. Gazeau, and D. Noguera, \textit{in preparation}.

\bibitem{bercugaro13} H. Bergeron, E.~M.~F. Curado, J.-P. Gazeau, and Ligia M.~ C.~ S. Rodrigues, Symmetric generalized binomial distributions, \textit{J. Math. Phys.}  \textbf{54} (2013) 123301-1-22 (2012).


\bibitem{cugaro10} E.~M.~F. Curado, J.-P. Gazeau, and Ligia M.~ C.~ S. Rodrigues, Nonlinear coherent states for optimizing quantum information, \textit{Phys. Scr.} \textbf{82} (2010) 038108-1-9.

\bibitem{cugaro11} E.~M.~F. Curado, J.-P. Gazeau, and Ligia M.~ C.~ S. Rodrigues, On a Generalization of the Binomial Distribution and Its Poisson-like Limit, \textit{J. Stat. Phys.} \textbf{146} (2012)  264-280 .

\bibitem{bercugaro12} H. Bergeron, E.~M.~F. Curado, J.-P. Gazeau, and Ligia M.~ C.~ S. Rodrigues, Generating functions for generalized binomial distributions, \textit{J. Math. Phys.}  \textbf{53} (2012) 103304-1-22 (2012).

\bibitem{mandel58} L. Mandel, Fluctuations of Photons Beams and their Correlations, \textit{Proc. Phys. Soc. (London)} \textbf{72}	(1958) 1037-1048; Fluctuations of photon beams: the distribution of photoelectrons, \textit{Proc. Phys. Soc.} \textbf{74}, (1959) 233-243. 
 
 \bibitem{mandelwolf70} L. Mandel and E. Wolf,  \textit{Selected Papers on Coherence and Fluctuations of Light}, vols. 1 and 2, Dover, New York, 1970.
 
 \bibitem{klyshko96}  D.~N. Klyshko, Observable signs of nonclassical light, \textit{Phys. Lett. A} \textbf{213} (1996) 7-15.  
 
\bibitem{koczyk96} P. Koczyk, P. Wiewior,  and C. Radzewicz, Photon counting statistics - undergraduate experiment, \textit{Am. J. Phys.} \textbf{64} (1996) 240-245.

\bibitem{loudon00} R. Loudon, The Quantum Theory of Light, 3rd ed., Oxford University Press (2000). 

\bibitem{gerry04} C. Gerry and  P. Knight, \textit{Introductory Quantum Optics}, Cambridge University Press (2004).

\bibitem{bachor04} H.~A. Bachor and T.~C. Ralph, \textit{a guide to experiments in Quantum Optics},  Wiley-VCH (2004).

\bibitem{eimigpol11} M. D. Eisaman, J. Fan, A. Migdall, and S. V. Polyakov, Single-photon sources and detectors (Invited Review Article), \textit{Review of Scientific Instruments} \textbf{82} (2011) 071101-25.

\bibitem{rodriguezlara17} C. Huerta Alderete, Liliana Villanueva Vergara, and B.~M. Rodr\'{\i}guez-Lara, Nonclassical and semiclassical para-Bose states, \textit{Phys. Rev. A}  \textbf{95} (2017)  043835-1-7. 

\bibitem{dirac27a} P.~A.~M. Dirac, The Quantum Theory of Emission and Absorption of Radiation, \textit{Proc. Royal Soc. Lond. A} \textbf{114} (1927) 243-265.  




%\bibitem{koelinkjeugt97} H.~T. Koelink and  J. Van der Jeugt, Bilinear generating functions for orthogonal polynomials,
	%arXiv:q-alg/9704016
	


%\bibitem{akhglaz81} N. I. Akhiezer and I. M. Glazman, \textit{Theory of Linear Operators in Hilbert Space} (Pitman, 1981).
%
%\bibitem{abraste72} M. Abramowitz and I.A. Stegun (Eds.),  \textit{Handbook of Mathematical Functions with Formulas, Graphs, and Mathematical Tables}, 9th printing (Dover, New York, 1972).




\end{thebibliography}
\end{document}